\documentclass[12pt]{article}

\usepackage{amsmath,amssymb,amsfonts,mathrsfs,mathtools,dsfont,mathrsfs, amsthm, bm, bbm}
\DeclareMathAlphabet{\mathpzc}{OT1}{pzc}{m}{it}
\DeclareSymbolFontAlphabet{\amsmathbb}{AMSb}%
\DeclareMathOperator*{\argsup}{arg\,sup}

\usepackage{bm,bbm,dsfont,epsfig,rotating,setspace,latexsym,amsmath,epsf,amsthm,amssymb,amsfonts,epstopdf,graphicx, cancel}

\usepackage{natbib}
 \bibpunct[, ]{(}{)}{,}{a}{}{,}%

\usepackage{cite,authblk}
\usepackage[dvipsnames]{xcolor}

\usepackage[normalem]{ulem}
\usepackage{comment}

\usepackage[dvipsnames]{xcolor}
\usepackage[font=small]{caption}
\usepackage[caption=false, font=small]{subfig}
\usepackage[hyphens]{url}
\usepackage[breaklinks, colorlinks, linkcolor=MidnightBlue, anchorcolor=MidnightBlue, citecolor=MidnightBlue, urlcolor=MidnightBlue]{hyperref}
\usepackage{subcaption}

\usepackage{array}

\usepackage[dvipsnames]{xcolor}
\usepackage{enumerate,enumitem}
\usepackage{csvsimple,booktabs}
\usepackage{float,subfig}
\usepackage{graphicx,epsfig,xcolor}
\usepackage{multirow}

\usepackage{fullpage}

\usepackage{ebgaramond}
\usepackage[OT1]{fontenc}
\usepackage[utf8]{inputenc}

\usepackage{thmtools}
\declaretheorem{theorem}
\declaretheorem{proposition}

\title{Pricing Problems in Adoption of New Technologies}

\author{Yijin Wang \qquad Subhonmesh Bose
\thanks{All authors are affiliated with the Department of Electrical and Computer Engineering and the Coordinated Science Laboratory at the University of Illinois Urbana-Champaign, Urbana, IL 61801. Emails: \{yijinw3, boses\}@illinois.edu. This work was partially supported by the grant NSF CAREER 2048065 and NSF ECCS 2349418 from the U.S. National Science Foundation.}}

\begin{document}

\maketitle

\begin{abstract}
We propose a generalization of the Bass diffusion model in discrete-time that explicitly models the effect of price in adoption. Our model is different from earlier price-incorporated models and fits well to adoption data for various products. We then utilize this model to study two decision-making problems. First, we provide a series of structural results on optimal pricing strategies to maximize profits from product sales by a monopolist over a finite horizon. We fully characterize the optimal pricing strategy in the single-period problem, and establish several structural properties of the same for the multi-period counterpart. 
Second, we study a Stackelberg game between a policy-maker and a monopolist, where the former seeks to maximize adoption through rebates, while the latter focuses on profits. For this problem, we analytically characterize crucial properties of the equilibrium path of the single-period game, and demonstrate how they carry over to the multi-period variant.
\end{abstract}

\section{Introduction}
\label{sec:intro}
The product diffusion model by \citet{bass} describes the aggregate first-purchase growth of a durable good in a market through time. Elegant in its simplicity, a differential equation encapsulates the effect of early adopters and imitators in the sales and life-cycle of a new product. Its predicted S-curves draws on \emph{diffusion of innovations} theory that owes its empirical foundations in \citet{ryan1950acceptance} from the analysis of the adoption of hybrid corn seed in Iowa and the later theoretical framework delineated in \citet{diffusion_innovations}. The model has been successfully applied across a plethora of markets, including consumer durables, electronics, pharmaceuticals (medical devices, new drugs), automotive (electric vehicles), agriculture (new hybrid seeds and fertilizers), enterprise software, and telecommunications among others. \citet{sultan1990diffusion} provides an early survey. Today's software platforms for modeling complex systems such as \citet{anylogic} with applications to healthcare and transportation leverages such a model internally. At its core, the Bass model captures the impact of social spillovers or peer effects in product diffusion. Collectively, adoption of new products is impacted by network externalities and often carries social signals, the collective impact of which is a positive feedback loop in sales. See \citet{peres2010innovation} for a deep dive into the threads of such influences.

Marketing strategies such as pricing and advertising obviously influence product growth--an element conspicuously missing in the early Bass model. In essence, the model endogenizes the effect of such strategic variables in its parameters as long as they remain constant over time or they change by a constant fraction over time. Optimal pricing strategies with price-driven product sales models emerged in the 1980's and 1990's, notable examples being \citet{bass1982note},  \citet{kalish1983monopolist}, \citet{kalish1983subsidy}, \citet{thompson1984oligopoly}, \citet{dockner1988oligopoly}, and \citet{bayus1992dynamic}. The aforementioned sources study pricing questions from a variety of angles, including impacts of advertising, peer effects, competition in oligopolies, and subsidies on product diffusion. Price-augmented Bass models became popular too; see e.g. \citet{dynamic_price_model}, \citet{mahajan1990new}, \citet{krishnan1999optimal}. \citet{krishnan1999optimal} derive optimal price paths using the \emph{generalized Bass model (GBM)} from \citet{gbm}. In this paper, we propose an alternate price-augmented product diffusion model that fits data from several markets and is amenable to mathematical analysis. We use the model to obtain structural insights into optimal pricing of a monopolist and then formulate and analyze a game between a rebate designer and a monopolist.

\citet{hbr_1976} underscores how price dynamics play a vital role in product adoption paths. 
% Indeed, one of our primary goals is to understand how pricing affects adoption dynamics. There have been a number of extensions to the Bass model which incorporate prices. \citet{dynamic_price_model} modified the Bass model in order to obtain dynamic pricing models. 
% \citet{gbm} later developed the Generalized Bass Model (GBM), a variant of the Bass model which introduces a pricing factor, and which reduces to the Bass model when the price is constant. 
While \citet{krishnan1999optimal} and \citet{krishnan2006} use the GBM to study optimal pricing/advertising, \citet{utility} points out several problems with the model itself and the resulting parameter estimation process with such a model. These models leverage a combination of absolute prices and relative price (and/or advertising) changes in their models. \citet{utility} deem dependency of product diffusion speeds on such a mix to be `untested'. Moreover, sales path predictions are sensitive to choices of parameter  initializations, as \citet{vandenbulte} finds out. 
% \citet{song_chintagunta_forward_looking_consumers} and \citet{melkinov_forward_looking_consumers} estimate utility-based models of adoption while assuming that consumers are forward-looking and thus solve an optimal stopping time problem to decide when to adopt. Nevertheless, they do not study optimal pricing from the firm's perspective, which is what we intend to do.
In Section \ref{sec:adoption model}, we introduce a new  adoption model in discrete-time that depends on prices and peer effects. In line with \citet{utility}, we show that our diffusion model can be obtained by allowing  individual agents to optimize their utilities in a large population. Our model departs significantly from prior models in that regardless of parameter choices or prices chosen at various times, mathematically \emph{the fraction of adopters $F_t$  remains strictly within $[0,1]$}. Specifically, we assume that the hazard function of the fraction of adopters follows a \emph{logit model} that grows with the innovation coefficient $p$ and imitation coefficient $q$, but reduces with price in that period. The model fits well to adoption datasets, and in fact outperforms both the Bass model and the GBM using certain measures of fit as reported in \citet{gbm}. In addition to fitting well to data, it is much more amenable to mathematical analysis. In particular, in Section \ref{sec:optimal pricing problem}, we formulate an optimal pricing problem over a fixed horizon $T$ with the adoption dynamics of Section \ref{sec:adoption model} and derive a closed-form optimal pricing strategy for the single-period problem ($T=1$). For the multi-period problem, we establish several structural properties of the maximum profit-to-go function and characterize monotonicity properties of optimal pricing when the coefficient capturing peer effects $q$ is small ($\leq 1$). These analyses allow us to simulate and study optimal price paths in Section \ref{sec:sales.path} that show features similar to those obtained in \citet{price_path} and \citet{kalish1983monopolist}. Such analyses have largely proven difficult with aforementioned price-augmented extensions of the Bass model. From a managerial standpoint, our derivation of the closed-form expressions for the $T=1$ problem provides valuable insights into the various trade-offs in optimal pricing and how they vary with the parameters. Furthermore, the structural results for the case with $T>1$ reveals how profits-to-go and optimal prices should behave with time and more adoption, and how these may qualitatively differ with high and low peer effects.

In the earliest stages of introduction of a new product into a market, there is often a lack of immediate willingness to buy it, as consumers may delay adoption in anticipation of future price drops and product improvements. Government promotion of new products can align with broader social goals, for example, with the adoption of green technologies such as rooftop photovoltaics (PVs) and electric vehicles (EVs). Large-scale adoption of PVs and EVs are widely recognized as meaningful  pathways for decarbonization and hence, define mechanisms to combat climate change. For these products, policymakers including local, state, and federal governments are motivated to catalyze broader diffusion. \citet{ev_prices_incentives} identified purchase prices as the most important factor in a customer's EV purchase decision. Policymakers can directly affect a customer's out-of-pocket payments for purchase through financial incentives.
% to consumers when adopting a product that the government deems to be beneficial to the public, environment, economy, etc. 

Rebate as a policy instrument to aid product diffusion has been widely recognized. For example, in 2022, the U.S. federal government began offering the Residential Clean Energy Tax Credit which gives homeowners a 30\% tax credit for installation of clean energy products such as rooftop PV systems \citep{percentage_tax_credit}, valid up to 2032. Qualifying new EVs (acquired before September 2025) have been eligible for a federal tax credit up to \$7,500. Even certain used EVs could qualify for such a credit up to \$4,000, per the U.S. Internal Revenue Service. Home EV charging equipment installations can provide credits up to \$1,000 through July 2026. Many states, including Colorado, California, New York, and Illinois, provide additional rebates between \$500 and \$5,000. See 
\citet{electrek2025stateev} for a survey of such credits. 

Rebates for PV systems and EVs are not confined to the U.S., but are common elsewhere. To provide some examples, Germany provides a 10-year vehicle tax exemption for battery-powered EVs registered before 2026, with additional reductions for cars priced less than \texteuro 70,000. France grants up to \texteuro 4,000 for new battery-operated EVs with further subsidies for charging infrastructure installation. 
Spain allows 15\% income tax deduction based on the purchase price, up to \texteuro 3,000, with a subsidy between \texteuro 4,500-7,000 for replacing old cars with battery-powered EVs. For a more detailed list on Europe's EV subsidies, see \citet{vecharged2025europeev}. Canada's Greener Homes Loan program grants interest-free loans up to CAD 40,000 for homeowners to accomplish energy-efficient retrofits, including installation of PV systems. Even Canadian businesses are eligible up to 30\% refundable tax credit for investing in commercial or industrial solar PV systems, according to \citet{ca_investment_tax_credit}. Prominent rebates in Asia include up to \textyen 22,500 ($\approx$ \$3,500) for EVs with driving ranges above 400km and up to INR 2 lakh ($\approx$ \$2,400) for EVs, as pointed out by \citet{ev_by_country}.

For policymakers, targeting clusters of early adopters through financial incentives can aid initial sales that in turn can spur further adoption through both financial incentives and peer effects, per \citet{peer_effects}. Adoption of PV systems and EVs have been examined for peer effects, including those via Bass-type diffusion models, e.g., see \citet{xia2022evadoption, pathania2017solar, ddcm, wang2024diffusion, zapata2024diffusion, yu2016market}, among others. In this work, we are interested in studying the interplay of rebate design and monopolist pricing in product adoption dynamics. Notice that rebates are often offered as a fixed sum (e.g., rebates for battery-operated EVs in France) or as a percentage of costs faced by consumers (e.g., residential clean energy tax credit from the U.S. federal government). In this paper, we adopt the former rebate design type and leave the analysis of the latter for future work.

% Note that the  is offered as a percentage of costs, whereas the EV tax credit is a fixed sum. Even though the former strategy seems more enduring and requires little adjustment due to inflation (after all, the fixed tax credit of 30\% of costs is expected to remain in effect for a decade since its was introduced), we will instead adopt the latter strategy of fixed absolute rebates in our work, as it is more amenable to mathematical analysis. 

In Section \ref{sec:game}, we analyze a Stackelberg game between a firm selling a clean energy product and a policymaker who wants to encourage adoption of the product through rebates. The game's dynamics are almost the same as those which govern the aforementioned optimal pricing problem, except with the addition of rebates. The firm, which is the follower, must determine its optimal pricing strategy in response to a fixed rebate set by the leader, the policymaker. We analytically characterize the Stackelberg equilibrium in the single-period game: we show that the optimal rebate is unique, {and provide a practical range of problem parameters such that the optimal rebate is both nonzero and does not exceed the price (otherwise, customers would effectively be paid to adopt). We also show that} the presence of rebates does indeed boost adoption, and that the optimal rebate decreases with increasing tendency of the population to adopt (i.e. higher values of the coefficients of innovation and imitation). 

For the multi-period game, we conduct numerical simulations and show that characteristics of the single-period game carry over to the multi-period counterpart. {Furthermore, our simulations suggest that with longer time horizons, the optimal rebate decreases.}
We remark that while \citet{solar_rebate} and \citet{kalish1983subsidy} provide insight into effective rebate design, we believe that our game-theoretic perspective on optimal rebate design and pricing strategies stands as a novel contribution. The game formulation is motivated by the fact that while  
green technology companies by nature seek to maximize profit and not the total number of adopters, policymakers by design have a different goal, and hence, must 
aid to steer the market towards the social goal, nevertheless allowing market forces to adjust to their steering actions.

Concretely, the contributions of this work are as follows.
\begin{itemize}
    \item We provide a new price-adjusted product adoption dynamics in discrete-time that fits adoption data well and aids analysis of pricing problems with said underlying dynamics.

    \item We analytically characterize several structural properties for the monopolist pricing problem and simulate its optimal pricing paths.

    \item We formulate and study a game-theoretic formulation of the question of rebate design between a policymaker and a monopolist and establish several policy-relevant properties of the optimal rebate.
\end{itemize}
We believe that our work will inform the price and rebate-related decisions of  firms and policymakers, respectively, especially for green energy technologies such as EVs and PV systems.

\section{The Product Adoption Model} \label{sec:adoption model}
The Bass diffusion model in \citet{bass} describes how the proportion of adopters $F(t)$ of a new product in a fixed population  evolves in continuous time,
\begin{equation}
    \frac{dF(t)}{dt} = (1-F(t))(p + qF(t)). \label{eq:bass}
\end{equation}
The rate of adoption among the non-adopters at time $t$ depends on $p+q F(t)$ for non-negative parameters $p, q$ that encode the propensity for innovation in adoption and imitation of other adopters, respectively. For $p < q$, the growth of $F(t)$ follows an S-shaped curve, where innovation drives slow growth early on, but grows rapidly later, driven by peer effects, before saturating at unity.

\begin{figure}[h!]
\centering
  \includegraphics[scale = 0.43]{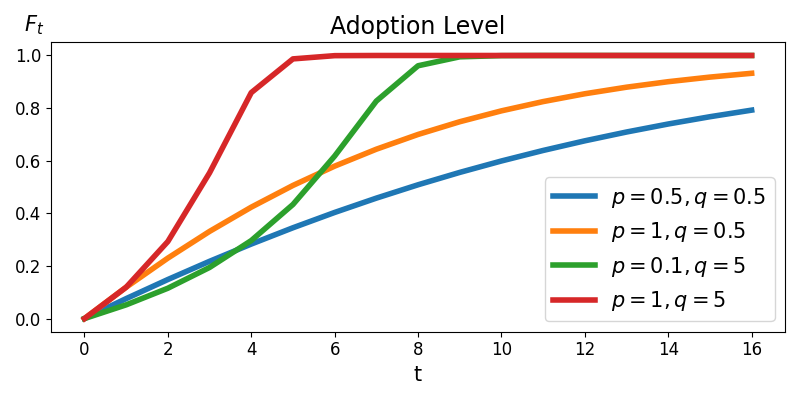}
  \caption{Trajectory of $F_t$ for a fixed price of $\pi = 3$, with $\alpha = 1$.}
  \label{fig:fixed price}
\end{figure}

We develop a diffusion model that includes the effect of price in the adoption dynamics, similar in spirit to GBM in \citet{gbm} and its variants in \citet{krishnan1999optimal}, but with a crucial difference as we describe next. One of the models in \citet{krishnan1999optimal} is given by
\begin{equation}
\frac{dF(t)}{dt} = (1-F(t))(p + qF(t)) \times \left(1 + \gamma \ln\left(\pi(0)\right) + \beta\frac{\frac{d\pi(t)}{dt}}{\pi(t)} \right) \;,
\label{eq:gbm with k}
\end{equation}
where $\pi(t)$ is the price at time $t$, $\eta$ (generally negative) is the sensitivity of adoption rate to changes in price, and $\gamma \leq 0$ measures the sensitivity to the initial price. A crucial shortcoming of this model (among others) is that the product multiplying $(1-F(t))(p + qF(t))$ on the right-hand side of the above relation can make $F(t)$ leave the interval $[0,1]$. Explicitly forcing that to remain within that interval renders ensuing analysis of optimal pricing problems challenging. In our model, we circumvent this challenge via a logistic function based factor that encodes the influence of price in the following discrete-time dynamics,
\begin{equation}
    F_{t+1} = F_t + (1-F_t) \frac{e^{p + qF_t - \alpha\pi_t}}{1 + e^{p + qF - \alpha\pi_t}}\;,
    \label{eq:F+ complete} 
\end{equation}
where $\alpha > 0$. For sufficiently small ${p}/{q}$ and a fixed price $\pi$, the trajectory of adoption mimics the Bass-type S-curves as Figure \ref{fig:fixed price} reveals. By design, $F_t \in [0,1]$ with $F_0 \in [0,1]$, regardless of the price---a feature that is oddly absent in the vast majority of the models proposed in the literature. Next, we demonstrate that our model in \eqref{eq:F+ complete} can be explained as the evolution of the fraction of adopters in a large population of agents where each agent is endowed with a specific utility function, akin to that developed in \citet{utility} for the Bass-logit model. 
% shows the dynamics of \eqref{eq:F+} for different $p$ and $q$ and a fixed price. The two scenarios in which $q = 5$ both exhibit the behavior of the ``S" curve.  

%-----

\subsection{Adoption Dynamics Via Agent Utilities}
Consider a population of $N$ agents. Suppose agent $i$ has not adopted till time $t$. Let her utility from adoption be given by
\begin{align}
U_t^i = p + qF_t - \alpha\pi_t + \varepsilon^i_t,
\end{align}
where $\alpha > 0$ and $\varepsilon^i_{t}$ is a random shock. Assume $\varepsilon^i_{t} \sim \text{Logistic}(0, 1)$, with cumulative distribution function $P(\varepsilon^i_{t} \leq x) = \frac{1}{1 + e^{-x}}$, and are independent and identically distributed for all $i,t$. Agent $i$ adopts at time $t$ if $U_t^i \geq 0$. Then, 
\begin{equation}
    P\left(U_t^i > 0\right) 
     = \frac{e^{p + qF_t - \alpha\pi_t}}{1 + e^{p + qF_t - \alpha\pi_t}}.
    \label{eq:prob.U}
\end{equation}
Letting $A_t$ denote the set of adopters till time $t$ with cardinality $|A_t| = N F_t$, we deduce
\begin{equation}
N F_{t+1} 
= N F_t + \sum_{i \notin A_t}\boldsymbol{1}\{U_{i, t} > 0\},
\label{eq:N t+1}
\end{equation}
with $\boldsymbol{1}$ as the indicator function. 
In turn, this implies
\begin{subequations}
    \begin{align}
    F_{t+1} 
    &= F_t +  \frac{N-|A_t|}{N} \left( \frac{1}{N-|A_t|} \sum_{i \notin A_t}\boldsymbol{1}\{U_{i, t} > 0\} \right)
    \\
    &\to F_t + (1-F_t) P(U^i_t > 0)
    \label{eq:F.prob}
\end{align}
\end{subequations}
almost surely, owing to the strong law of large numbers as $N \to \infty$. Replacing \eqref{eq:prob.U} in \eqref{eq:F.prob} gives our price-adjusted diffusion model in \eqref{eq:F+ complete}.

%-----

\subsection{Model Evaluation Using Data}
In order to evaluate the practicality of our model, we fit historical adoption and pricing data for new products. We used the same datasets for the three products on which the GBM \citep{gbm} was evaluated: air conditioners, color TV, and dryers for clothes. Furthermore, to see how our model performed on more recent data, we fit it to U.S. residential solar PV pricing and adoption data through 2023 from \citet{solar_data}. The model performed generally favorably on the GBM household appliance datasets, but we suspect signs of systematic bias in fitting the model to solar data. 

The pricing and sales data for the household appliance products were taken directly from Table 7 in \citet{gbm}. The adoption level at year $t$, $F_t$, was calculated by dividing the total sales till year $t$ by the estimated population size for each product, which were taken from Table 3 in \citet{gbm}. 
The solar data \citep{solar_data} includes historical residential solar installations, along with their capacities, in kilowatts, as well as total installed prices, in nominal U.S. dollars. The adjusted prices for each year were calculated by taking the median of the per-Watt prices of the installations in that year, and converting it to 2024 dollars using a Consumer Price Index calculator \citep{inflation_calculator}. 
%\bose{Full form of CPI.} 
The solar installation data was filtered for each state, and the adoption level at year $t$, $F_t$, was calculated by dividing the total installations up till year $t$ by the 2023 population size for each state, obtained from \citet{census}. 

We estimated the model parameters using nonlinear model identification functions from Matlab's System Identification Toolbox. Specifically, we used the nonlinear grey-box model toolbox which estimated the model parameters using nonlinear least-squares. The algorithms output the normalized root mean squared error (NRMSE) for each fit as
\begin{equation}
    \text{NRMSE} = \frac{\left\Vert y_{\text{data}} - y_{\text{model}}\right\Vert_2}{\left\Vert y_{\text{data}} - \overline{y}_{\text{data}}\right\Vert_2} \;, \label{eq:nrmse}
\end{equation}
where $y_{\text{data}}$ are the adoption level data $F_t$ obtained from the datasets using the aforementioned methods, $\overline{y}_{\text{data}}$ is the average of these data, and $y_{\text{model}}$ are the adoption levels estimated by our model. 
%\bose{What is the one with the bar above? Define it. Use the notation $\langle y_{\textrm{data}} \rangle$.}
 
\begin{figure}[!h]
\centering
\includegraphics[scale=0.58]{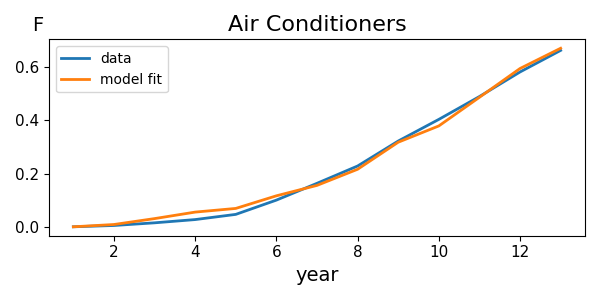}
\vspace{0.1cm}
% % \raggedright
\includegraphics[scale=0.58]{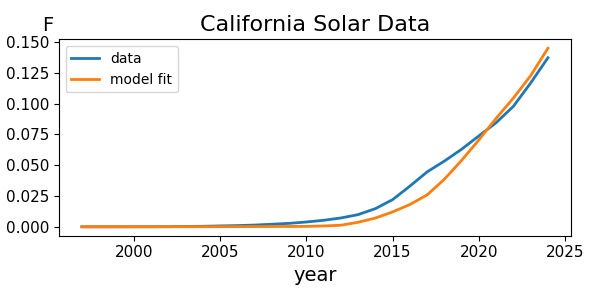}
\caption{Model fit for GBM air conditioner data (top) and California solar prices (bottom).}
\label{fig:model fit}
\end{figure}

Table \ref{tab:fit results} reports statistics of the model fit for each dataset. The results for the solar PV data for various states were similar, so we only chose to present the results for California. The $R^2$ values are calculated as $1 - \text{NRMSE}^2$, where NRMSE is defined in \eqref{eq:nrmse}. The $R^2$ values are within an acceptable range and out model fit outperforms those for the GBM and the original Bass model on the first three datasets, per Table 3 in \citet{gbm}. The fits on air conditioner and California residential solar PV data are shown in Figure \ref{fig:model fit}. As is the case for all the states' solar data we fit the model on, our model consistently under- then over- estimates the true data, as evident in Figure \ref{fig:model fit}; we suspect some model bias. 
\begin{table}[h!]
\caption{Model Fit Performance Results}
\label{tab:fit results}
\centering
\begin{tabular}{lrr}
\toprule
 & NRMSE & $R^2$ \\
\midrule
Air Conditioners & 0.0661 & 0.9956 \\
Television & 0.0650 & 0.9958 \\
Dryers & 0.1243 & 0.9845 \\
California Solar Data & 0.1727 & 0.9702 \\
\bottomrule
\end{tabular}
\end{table}

% Despite the possibly poor fit for some datasets, we designed our model to be, first and foremost, 
While the results are encouraging, the model derives its utility in being amenable to mathematical analysis, as the next few sections will reveal.

%-------

\section{The Optimal Pricing Problem}
\label{sec:optimal pricing problem}

Consider a monopolist that seeks to maximize its profit by setting prices at times $t=0, \ldots, T-1$. Higher prices in a time step reduce adoption at that time step and negatively affects the adoption growth through peer effects at later periods, even though it yields higher profits per fraction of the population that adopts in that time step. In this section, we formulate and analyze the finite-horizon deterministic optimal control problem for profit maximization $\mathcal{P}$ with our adoption model in \eqref{eq:F+ complete}. A quick comparison of this analysis with those in \citet{dynamic_price_model}, \citet{bass1982note}, \citet{kalish1983monopolist}, and \citet{kalish1983subsidy}, etc. reveals the analytical tractability of our model.

Setting the price $\pi_t$ at time $t$, let the profit be $\pi_t - C$ per unit fraction of the population that adopts the product. Here, $C \geq 0$ describes a constant cost of production.\footnote{An alternate model is to account for decreasing costs to the manufacturer over time with maturing technology. The conclusions from our analysis can be easily extended to account for the same.} Recall from \eqref{eq:prob.U} and \eqref{eq:F.prob} that setting a price $\pi_t$ leads to the adoption by a fraction $(1-F_t)R(F_t, \pi_t)$ of the population, where
\begin{subequations} \label{eq:B and R}
\begin{equation}
    B(F, \pi) \coloneqq e^{p + qF - \pi}
  \label{eq:B}  
  \end{equation}
  \begin{equation}  
    R(F, \pi) \coloneqq \frac{B(F, \pi)}{1 + B(F, \pi)} 
    % = \frac{e^{p + qF - \pi}}{1 + e^{p + qF - \pi}} \;. 
    \label{eq:R}
\end{equation} 
\end{subequations}
For the remainder of the paper, we will assume $\alpha=1$ with the understanding that our results can be easily extended to the case with $\alpha \neq 1$. With this notation, the profit maximization problem $\mathcal{P}$ is
\begin{equation}
    V(F_0) \coloneqq \sup_{\pi_0, \ldots, \pi_{T-1}\geq 0}\sum_{t = 0}^{T-1} (\pi_t - C)(1-F_t) R(F_t, \pi_t) \;, \label{eq:P}  
\end{equation}
where $V$ is the optimal profit function given the initial proportion of adoption $F_0$. We seek a pricing policy $\left( \pi_0, \ldots, \pi_{T-1} \right)$ that at each time $t$ maps $F_t$ to a non-negative price, starting from $F_0$. The total profit over a horizon of length $T$ is $V(F_0)$.

Several comments are in order about the nature of our formulation of $\mathcal{P}$. First, our formulation is over a finite horizon $T$, where this horizon can be viewed as the monopolist's expectation of how long its market for the first-purchase of the new product will run. In general, $T$ is unknown to a firm that seeks to solve $\mathcal{P}$, but firms often have expectations for how long they expect a product to last in the market before disruptive factors such as a completely different technology alters the market dynamics significantly. One can alternately model this expectation through an infinite horizon formulation with future cash-flows discounted at an exponentially decaying rate--a route that can be explored with our diffusion dynamics in a future work. Second, our pricing problem is a deterministic optimal control problem, given that our adoption dynamics in \eqref{eq:F+ complete} is deterministic. While previously studied diffusion models are largely deterministic in nature, we acknowledge that firms often face a range of uncertainties in launching a new product, e.g., parameters such as $p$ and $q$ are not known a priori and can be challenging to estimate at the take-off stage, as \citet{mitra2019forecasting} points out. 
% The ensuing analysis in the sequel with our deterministic model as a first step towards one that accounts for said uncertainties.

\subsection{Towards Solving $\mathcal{P}$}

We leverage dynamic programming from  \citet{dp} and define the profit-to-go function at time $t$ recursively as 
\begin{equation}
\begin{aligned}
    V_{T-1}(F_{T-1}) 
    \coloneqq
    & \sup_{\pi_{T-1} \geq 0}\left(\pi_{T-1} - C\right)\left(1-F_{T-1}\right) R(F_{T-1}, \pi_{T-1}),
    \\
    V_t(F_t) 
    \coloneqq & \sup_{\pi_t \geq 0} \left\{ (\pi_t - C)(1-F_t) R(F_t, \pi_t) + V_{t+1}\left(F_t + (1-F_t) R(F_t, \pi_t)\right) \right\}, \\
    & \qquad \qquad \qquad \forall t = 0, \ldots, T-2 \,.
\end{aligned} \label{eq:dp}
\end{equation}
By the dynamic programming principle, $V(F_0) = V_0(F_0)$. We now analyze the profit-to-go functions $V_0, \ldots, V_{T-1}$, backward in time.
% The optimal pricing policy $\pi_t^\star(F_t)$ is such that the objective in $V_t(F_t)$ achieves its supremum.
%-------------------------
% \subsection{Analysis of the Optimal Profit-To-Go Functions} 
% \label{sec:opt.val}
% The final price, $\pi_{T-1}$, governs the final proportion of adopters $F_T$. 
% We dedicate this section to understanding properties of $V_0, \ldots, V_{T-1}$ that shed light into the solution of the optimal control problem $\mathcal{P}$. 
We start by providing a closed-form solution for the optimal price at $t = T-1$, which is equivalent to solving the single-period optimal pricing problem with $T=1$.
\begin{theorem} \label{thm:pi* T-1}
The optimal price at $t = T-1$ is 
\begin{equation}
    \pi^\star_{T-1}(F_{T-1}) = C + 1 + W\left(e^{p + q F_{T-1} - (C+1)}\right) > 0\;, \label{eq:pi* T-1}
\end{equation}
    where $W$ is the principal branch of the Lambert-W function.
\end{theorem}
\proof{Proof.}
Define the incremental profit at each time as 
\begin{equation}
    H(F, \pi) \coloneqq (\pi-C)(1-F)R(F, \pi) \;. \label{eq:H}
\end{equation} 
Finding the optimal price at $t = T-1$ amounts to maximizing $H$ over $\pi$. Notice that
\begin{equation} \label{eq:lim H infty}
\lim_{\pi \rightarrow \infty} H(F, \pi)
% & = (1-F)\lim_{\pi \rightarrow \infty} \frac{\pi-C}{\frac{1}{R(F, \pi)}} \\
% & 
= \frac{1-F}{\lim_{\pi \rightarrow \infty} \frac{\partial}{\partial \pi}\left(\frac{1}{R(F, \pi)}\right)} 
% \label{eq:l'hopitals} \\
% & 
= 0
\end{equation}
due to l'H\^{o}pital's rule. 
Then, we have
\begin{equation}
\begin{aligned}
    \frac{\partial H(F, \pi)}{\partial \pi} 
    % &= (1-F)R(F, \pi)G(\pi) \\
    & = (1-F)R(F, \pi)\left(1 - \underbrace{\frac{\pi-C}{1 + e^{p + qF - \pi}}}_{\eqqcolon G_H(\pi)}\right).
\end{aligned} \label{eq:dHdpi}
\end{equation}
The numerator of $G_H$ strictly increases with $\pi$, while the denominator strictly decreases with $\pi$. Also, $G_H(0) \leq 0$ implies that the right-hand side of \eqref{eq:dHdpi} becomes zero at a unique value that satisfies
\begin{equation}
\begin{gathered}
 % \frac{\partial H(F, \pi^{\star})}{\partial \pi} = 0 \rightarrow G(\pi^{\star}) = 1 \\
    \pi^\star-C = 1 + e^{p + qF - \pi^\star},
\end{gathered}
    \label{eq:G pi *}
\end{equation}
Since $\partial H/\partial F$ is positive before $\pi^\star$ and negative post $\pi^\star$, we deduce that $\pi^\star$ defines the unique global maximum of $H(F, \pi)$ except for when $F = 1$, in which case $H(F, \pi) \equiv 0$.
The relation in \eqref{eq:G pi *} yields \eqref{eq:pi* T-1} using the properties of the Lambert-W function with $F_{T-1}$ in place of $F$.
%\Halmos
\endproof

Finding the optimal price at $t = T-1$ amounts to solving a single-period optimal pricing problem, where the optimal price extracts as much profit from adoption as possible in that one round. Higher price extracts more profit per unit fraction of adoption, but lowers the adoption rate. The optimal price balances these two factors. 

Our result permits qualitative insights into the nature of the optimal price. Notice that the Lambert-W function is increasing on $\mathbb{R}^+$, and so is $\pi^{\star}_{T-1}(F)$ in \eqref{eq:pi* T-1}, indicating that the greater the proportion of existing adopters, the higher the firm can afford to set its price. The optimal price is also larger with larger $p$ and $q$, since these factors reflect the willingness of consumers to adopt. Furthermore, since 
\begin{align*}
    \frac{\partial\pi^\star_{T-1}}{\partial C} = \left({1 + W\left(e^{p + q F_{T-1} - (C+1)}\right)}\right)^{-1} > 0 \;,
\end{align*}
the optimal price is increasing with production cost, which is expected, since the firm sets higher prices to compensate for higher costs. Interestingly though, $\pi^\star_{T-1}$ \emph{does not grow linearly with cost}, driven by the fact that the fraction of adopters depends on price in a nonlinear fashion.

The pricing problem at stages $t<T-1$ is not so straightforward as the prices set earlier in the time horizon affect the growth of fraction of adopters at later time periods, which then impact how much profit the firm can extract via pricing in the later periods. In other words, the greedy nature of the pricing policy in the last time-step may not be optimal for the multi-period pricing problem.
We now present a structural result on the optimal price and the profit-to-go functions for $\mathcal{P}$ with $T>1$. The proof is deferred to 
Appendix \ref{proof of prop:general}.
\begin{proposition}\label{prop:general} The search for the optimal prices at each time $t$ can be restricted to a compact set $[0, \overline{\pi}_t]$. Furthermore, $V_t$ is pointwise non-increasing with $t$, i.e., 
$V_t(F) \geq V_{t+1}(F)$ for $t = 0, \ldots, T-2$. 
\end{proposition}
Based on the above result, one can restrict the search for an optimal price at any time to a bounded interval. Such a result is expected, since an infinite price at any time would lead to zero new adopters (and zero profit) at that time. Such a policy can only be optimal, if gaining zero adopters remains optimal from then on, which can only happen if adopters have already saturated the market, i.e. $F_t = 1$. In other words, this result rules out that possibility in finite-time for maximizing cumulative profits. We also showed that the reward-to-go function is element-wise decreasing with time. This is  intuitive, because $V_t$ is the maximum profit earned over $t, \ldots, T-1$, so starting from later $t$ at the same adoption level, lesser time till the end translates to lesser (optimal) cumulative profits. 

%---------

\subsection{Optimal Pricing with Low Peer Effects} \label{sec:special case}
% In Propositions \ref{thm:pi* T-1} and \ref{prop:general}, we have already derived explicit expressions for the optimal pricing strategy and the reward-to-go function at $t = T-1$, but it remains challenging to find similarly closed-form solutions for $t < T-1$. 
The case with low peer effects ($q \leq 1$) permits the derivation of further structural insights that we present next; 
% Yet, when , i.e., when peer effects are low, we can show that the global optimal price is unique, positive, and a non-increasing function of the proportion of adoption. 
the proof is included in Appendix \ref{proof of thm:special case}.
\begin{theorem} \label{thm:special case}
Suppose $q \leq 1$. Then, the optimal price $\pi^{\star}_t(F)$ is unique for $F \in [0,1)$ and furthermore satisfies $\pi_t^{\star}(F) > 0$, $\left[\pi^{\star}_t\right]'(F) \geq 0, \forall t = 0,\ldots, T-1$, and $\pi^{\star}_t(F) \geq \pi^{\star}_{t+1}(F), \forall t = 0,\ldots, T-2$. 
\end{theorem}
{The above result 
suggests that at any given time, a higher adoption level results in a higher optimal price. With greater existing adoption, more consumers will likely adopt due to peer effects, and hence, the monopolist has no incentive to lower the price with increased demand.} The optimal price turns out to be unique, finite, and positive; combined with the fact that the profit-to-go function is differentiable, we can then deduce that the optimal price is guaranteed to be its stationary point, a fact that facilitates a numerical search for the optimal price.
    We have also shown in Theorem \ref{thm:special case} that the optimal price for a fixed adoption level decreases over time when $q\leq 1$.
A possible explanation for this phenomenon is that in earlier time periods, the firm has more opportunities to make a profit and to increase the adoption level, and so can afford to set higher prices compared to later time periods, for the same adoption level. In contrast, as $t$ approaches the time horizon, $T$, the firm sets lower prices to encourage a surge in adoption in the relatively short remaining time.

\begin{figure}[H]
\centering
\includegraphics[scale=0.55]{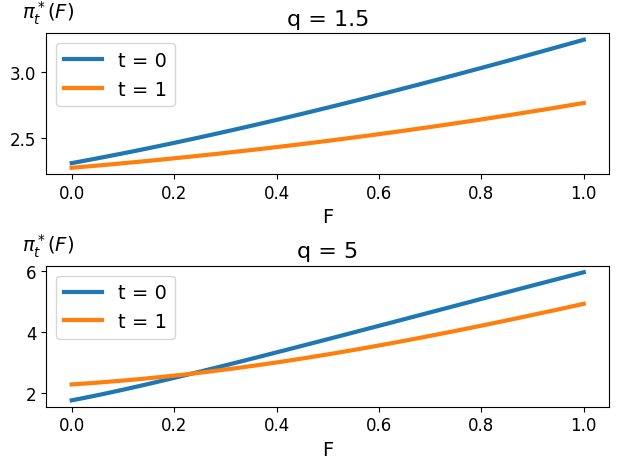}
\caption{Optimal Price for $T = 2$ for cases $q = 1.5$ (top) and $q = 5$ (bottom). The remaining parameters common to both cases are $p = 1, C = 1$.}
\label{fig:pi star}
\end{figure}

We now illustrate that the condition $q \leq 1$ is sufficient, but \emph{not necessary} for the monotonicity of optimal prices in Theorem \ref{thm:special case} to hold. Figure \ref{fig:pi star} displays simulated optimal pricing functions for $T = 2$ for two values of $q$; in both cases, $p = C = 1$. Note that for $q = 1.5$, which violates the condition in Theorem \ref{thm:special case}, we still have $\pi^{\star}_0(F) \geq \pi^{\star}_1(F), \forall F \in [0,1]$, i.e., the conclusion of the result holds, indicating that the condition $q \leq 1$ is not necessary. However, with $q = 5$, the conclusion breaks. 
%\bose{Find an explanation for why the ordering may not hold for higher values of $q$. }
A possible explanation is that for low values of $F$, and for smaller $t$, there is an incentive to lower the price so that the high peer effects can drive a substantial increase in adoption that increases price later in time. Such a possibility disappears when $q$ is low, where the optimal pricing strategy is predominantly dominated by remaining time considerations, in accordance with our earlier argument.

To perform the numerical simulations, we discretized the space of $F \in [0, 1]$ and populated the value function for each $F_t, t = 0, \ldots, T - 1$ using dynamic programming. 
We used the Optimization package in Python's Scipy library to find the optimal $\pi^\star_t(F_t)$ for each $F_t$. 
%All simulations were performed on an Intel\textregistered{} Xeon\textregistered{} Platinum 8272CL CPU @ 2.60GHz processor. \bose{Unless you are reporting solution times, such specs are not useful.}

\section{Optimal Price and Sales Paths}
\label{sec:sales.path}
% We used the same methodology as in Section \ref{sec:special case} to calculate the paths for the general case. 
The optimal control framework with our product diffusion model allows us to simulate optimal price and adoption paths, similar to those in \citet{dynamic_price_model}, \citet{bass1982note}, \citet{kalish1983monopolist}, \citet{price_path}, etc. 
We simulated said paths for different sets of parameters and varying time horizons using the same methodology as described in Section \ref{sec:special case}. 

\begin{figure}[!h]
\centering
\includegraphics[scale=0.6]{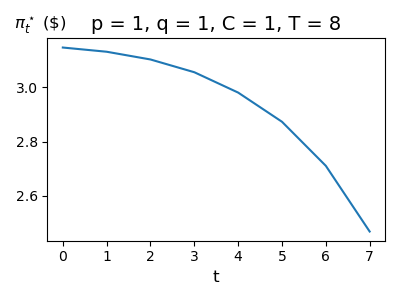}
\vspace{0.1cm}
% % \raggedright
\includegraphics[scale=0.6]{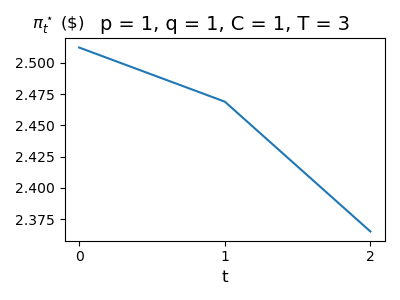}
\caption{Optimal pricing trajectories for $q = 1$ for different time periods.}
\label{fig:decreasing price}
\end{figure}

Our model exhibits behavior similar to those in existing results. For example, \citet{price_path} reports that when the peer effects are low, the optimal price monotonically decreases, whereas when peer effects are high, optimal price first increases and then decreases. Figure \ref{fig:decreasing price} displays the optimal price trajectory for when peer effects are low ($q = 1$) for two time horizons $T$; notice that for both $T = 8$ and $T = 3$, the optimal price monotonically decreases.

\begin{figure}[!h]
\centering
\includegraphics[scale=0.6]{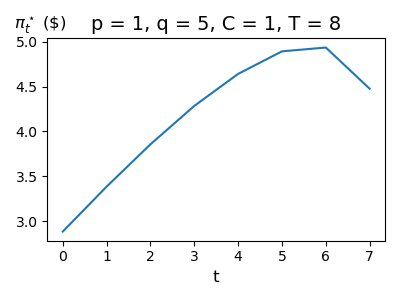}
\vspace{0.1cm}
% % \raggedright
\includegraphics[scale=0.6]{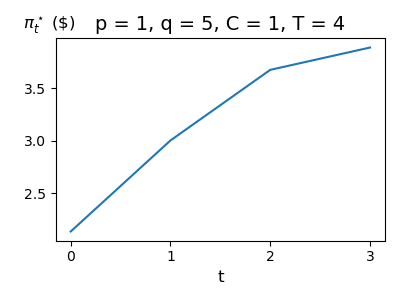}
\caption{Optimal pricing trajectories for $q = 5$ for different time periods.}
\label{fig:increasing decreasing price}
\end{figure}

\begin{figure}[!h]
\centering
\includegraphics[scale=0.6]{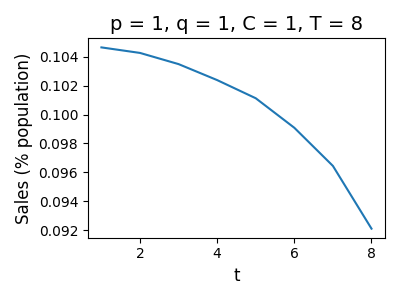}
\caption{Percentage of new adopters (as percentage of population) over time.}
\label{fig:sales}
\end{figure}

Figure \ref{fig:increasing decreasing price} shows that for high peer effects ($q = 5$), the optimal price path increases and then decreases when the time horizon is long enough (contrast the cases with $T = 4$ and $T=8$).
Our simulations are also in agreement with one of the results of \citet{kalish1983monopolist}; if the sales of new products decrease monotonically over time, the optimal price path also decreases monotonically. Figure \ref{fig:sales} illustrates this; compare this to the optimal pricing path of Figure \ref{fig:decreasing price} for the same set of parameters.

\begin{figure}[!h]
\centering
\includegraphics[scale=0.6]{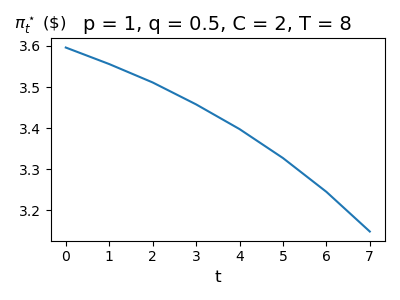}
\vspace{0.1cm}
% % \raggedright
\includegraphics[scale=0.6]{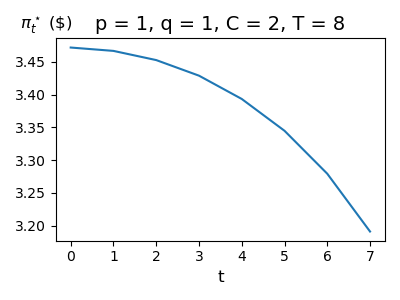}
\caption{Optimal pricing trajectories for 2 cases which differ only by the value of $q$.}
\label{fig:parameter change}
\end{figure}

Finally, the structure of the optimal pricing path is fairly insensitive to small shifts in parameters. For instance, Figure \ref{fig:parameter change} displays the optimal price path for two cases in which only the parameter $q$ varies slightly, yet the optimal price paths are both monotonically decreasing. Notice that the case $q = 1, C = 2$ only differs from the case in Figure \ref{fig:decreasing price} by the value of the parameter $C$; again, the shape of the optimal pricing path is similar. This suggests that our model is robust to some errors in the estimation of parameters. 

%------------

\section{Optimal Rebate Design via a Stackelberg Game} \label{sec:game}
%\bose{Rework the first two paragraphs to give examples of two ways of providing a rebate in the introduction. Here, start with how rebates can be effective and then plunge into this discussion below directly.}
Government incentives can be effective in spurring adoption of products with societal or environmental benefits, especially when the product is not yet widely adopted, and/or the costs to adopt are prohibitive, e.g., for PV systems and EVs. In the sequel, we will consider the question of rebate design along the lines of the Clean Vehicle Tax Credits that are offered as fixed sums. We consider the case where the policymaker first announces a fixed rebate that remains in effect for a period of time; meanwhile, a monopolist continuously adjusts their prices, taking into account the effects of the government-provided incentive on consumer willingness to adopt. This order of announcing the policies naturally lends itself to a Stackelberg game with the policymaker as the leader and the monopolist as the follower. In this section, we  formulate this game  and analyze its equilibria to glean insights into optimal rebate design and pricing responses from a monopolist. In general, multiple producers will vie for a share of the market that is influenced by the rebates from a policymaker. An analysis of rebate design for an oligopolistic competition is relegated to a future endeavor, for which our analysis provides the first step.

Consider a policymaker who sets a rebate, $r$, which is fixed for $t = 0, \ldots, T$, where $T$ defines a finite horizon. In response, the firm sets a price path $\pi = (\pi)_{t=0}^{T-1}$. The dynamics of adoption are almost identical to those of the optimal pricing problem in Section \ref{sec:optimal pricing problem}, except for the introduction of a rebate:
\begin{equation}
    F_{t+1} = F_t + (1-F_t)R\left(F_t, \pi_t, r\right) \;, \label{eq:F+ with r}
\end{equation}
where we use the notation,
\begin{equation}
\begin{gathered}
    B(F, \pi, r) \coloneqq e^{p + qF - (\pi - r)} \\
    R(F, \pi, r) \coloneqq \frac{B(F, \pi, r)}{1 + B(F, \pi, r)} \;. 
\end{gathered}\label{eq:B and R with r}
\end{equation}
In effect, the consumer faces a net price of $\pi - r$ to adopt the product when the firm responds with a price $\pi$ to the policymaker's announced rebate $r$. 
% It is possible that the consumer receives a rebate which exceeds the cost of the product, in which case the consumer is effectively paid money to adopt. For this scenario, we should expect a possible surge in adoption, which our model is ill-equipped to accurately describe. We will explore under what circumstances this case arises in Theorem \ref{thm:game}.

The policymaker seeks to maximize the fraction of adopters, while also minimizing the total rebate amount handed out, and hence, is modeled to maximize 
\begin{gather*}
V^p\left(F_0, \pi, r\right) := (F_T - F_0)\left(1 - \beta r\right) \;.
\end{gather*}
Here, $\beta > 0$ captures the relative emphasis that the policymaker places on reducing the rebate amount, compared to the growth in the fraction of adopters. 
% Let $V^p$ be the reward function for the policymaker, defined as 
Akin to the optimal pricing problem, the firm aims to maximize its own profits, which with a slight abuse of notation, is given by 
\begin{gather*}
V\left(F_0, \pi, r\right) = \sum_{t = 0}^{T-1} (\pi_t - C)(1-F)R\left(F_t, \pi_t, r\right),
\end{gather*}
with the announced rebate being $r$. 
For a given $r$, the follower (the firm) solves the optimal pricing problem,
\begin{equation}
\pi^{\star}(r) \in \argsup_{\pi = (\pi_0, \ldots, \pi_{T-1})\geq 0} V\left(F_0, \pi, r\right) \;. \label{eq:firm solution}
\end{equation}
This problem is almost identical to \eqref{eq:P}, except that $R(F_t, \pi_t)$ is replaced with $R\left(F_t, \pi_t, r\right)$. By comparing \eqref{eq:B and R} with \eqref{eq:B and R with r}, we deduce that the firm solves the optimal pricing problem from Section \ref{sec:optimal pricing problem}, where $p$ is effectively replaced by $p + r$. A strategy pair $\left(\pi^{\star}, r^{\star}\right)$ defines a Stackelberg equilibrium if $\pi^{\star}$ satisfies \eqref{eq:firm solution} for all $r \geq 0$ and 
\begin{equation}
r^{\star} \in \argsup_{r \geq 0} V^p\left(F_0, \pi^{\star}(r), r\right) \; . \label{eq:leader stackelberg}
\end{equation}

The game formulation between a policymaker and a monopolist highlights the competing nature of the goals of the two parties involved. The policymaker's goal is to spur adoption with the least amount of rebates, while the firm seeks to maximize its own profits. In our game formulation, we assume that both players know the parameters of the adoption dynamics for the product. Such parameters are generally unknown or difficult to estimate early on in the adoption cycle, but a monopolist and a policymaker will likely have access to similar market research data, and hence, it is reasonable to assume as a first cut that they will operate with similar  model parameters. Since the optimal pricing from the monopolist turned out not to be that sensitive to errors in parameter estimation, we expect the same behavior for an optimal rebate design as well. 

%---------

\subsection{Optimal Rebate in a One-Stage Game}
We now establish structural properties of a Stackelberg equilibrium of the game between the policymaker and the firm for the single-period game with $T=1$. These properties of optimal rebate design  carry over to the $T>1$ case, as our numerical experiments in Section \ref{sec: numerical game} will reveal.

% \bose{Also, after reading your proof, it seems straightforward to come up with a clean condition (using $G(0), \beta$) under which $r^\star =0$, i.e., rebates are not useful. That is a nice structural result and you can interpret it.}
\begin{theorem}\label{thm:game}
For the single-period Stackelberg game, 
% the optimal price $\pi^{\star}(r)$ is unique and positive. Furthermore, 
the optimal rebate, $r^\star$, is unique. 
Let
\begin{equation}
    \beta_0 = \left(1 + W\left(e^{p + qF_0 - (C+1)}\right)\right)^{-2}\;, \label{eq:beta 0}
\end{equation}
where $W$ is the principal branch of the Lambert-W function. When $\beta < \beta_0$, $r^\star > 0$; otherwise, $r^\star = 0$.
Furthermore, let 
\begin{equation}
\hat{\beta} = \left(C + 1 + e^{p + qF_0} + \left(1 + e^{p + qF_0}\right)^2\right)^{-1} \;. \label{eq:beta hat}
\end{equation}
% Let $\pi^{\star}(r)$ be the optimal price for a given rebate $r$. 
Then, for $\beta > \hat{\beta}$, $\pi^\star(r^\star) > r^\star$; otherwise, $\pi^\star(r^\star) \leq r^\star$.
Finally, when $r^\star > 0$, the final adoption level is greater than in the absence of rebates (i.e. in the single-stage adoption problem).
\end{theorem}
%\bose{Can we remove the dependence of $\pi^\star$ on $F$? The whole game is defined for a given $F_0$ and that is implicit in the game definition itself. Also, it should be $F_0$ instead of $F$ everywhere. Can you make the change?}
\proof{Proof.}
The firm solves a single-period optimal pricing problem with $p$ replaced by $p + r$, so by \eqref{eq:pi* T-1},
\begin{equation}
    \pi^{\star}(r) = C + 1 + W\left(e^{p + qF_0 + r - (C+1)}\right) > 0\;. \label{eq:pi star firm}
\end{equation}
The policymaker's objective is $V^p \left(F_0, \pi^{\star}(r), r\right)= (1-F_0)R(F_0, \pi^{\star}(r), r)(1-\beta r)$. Since $V^p\left(1, \pi^{\star}(r), r\right) = 0$, we only analyze the case $F_0 < 1$. %\bose{Why $\equiv$? It should be equal to...}
\begin{equation}
\frac{dV^p\left(F_0, \pi^{\star}(r), r\right)}{dr} = (1-F_0)R(F_0, \pi^{\star}(r), r) \left(G(r) - \beta\right) \;, \label{eq:dWdr}
\end{equation}
where 
\begin{equation}
G(r) \coloneqq \frac{1-\beta r}{\left(1 + B\left(F_0, \pi^{\star}(r), r\right)\right)\left(1 + W\left(e^{p + qF_0 + r - (C+1)}\right)\right)}\;. \label{eq:G(r)}
\end{equation}
%\bose{Can we change $G_{V_p}$ to $G_p$?}
In order to simplify $G(r)$, first note from \eqref{eq:B and R with r} and \eqref{eq:pi star firm} that 
\begin{equation}
\begin{aligned}
B\left(F_0, \pi^{\star}(r), r\right) &= e^{p + qF_0 + r - \left(C + 1 + W\left(e^{p + qF_0 + r - (C+1)}\right)\right)} \\
& = W\left(e^{p + qF_0 + r - (C+1)}\right) \;,
\end{aligned} \label{eq:B pi star with r}
\end{equation}
where the second equality uses the fact that $e^x = W\left(e^x\right)e^{W(e^x)}$. From the definition of $G(r)$ in \eqref{eq:dWdr} and using \eqref{eq:B pi star with r},
\begin{equation}
G(r) = \frac{1-\beta r}{\left(1 + W\left(e^{p + qF_0 + r - (C+1)}\right)\right)^2} \;.
 \label{eq:G W}
\end{equation}
The numerator of $G(r)$ decreases monotonically to $-\infty$. Since $W\left(e^{p + qF_0 + r - (C+1)}\right)$ increases with $r$, $G(r)$'s denominator is positive and increases monotonically in $r$. Thus, $G(r)$ decreases monotonically to 0, after which it remains negative. 

Setting $\beta_0 = G(0)$ gives us \eqref{eq:beta 0}. From \eqref{eq:dWdr}, if $\beta \geq \beta_0$, $V^p\left(F_0, \pi^{\star}(r), r\right)$ decreases monotonically in $r \in \mathbb{R}^+$, implying $r^{\star} = 0$. If on the other hand, $\beta < \beta_0$, $G(r)$ decreases monotonically to $\beta$ at some $r^\star> 0$, then remains smaller than $\beta$ for all $r > r^\star$. Thus, $V^p$ increases until the unique maximum $r^\star > 0$ and then decreases. From \eqref{eq:dWdr}, we have
\begin{align*}
    \frac{d}{dr} V^p\left(F_0, \pi^{\star}(r^{\star}), r^{\star}\right) = 0 \rightarrow G(r^{\star}) = \beta \;.
\end{align*}
Thus, from \eqref{eq:G W}, the optimal rebate satisfies
\begin{equation}
    \frac{1}{\beta} = r^{\star} + \left(1 + W\left(e^{p + qF_0 + r^{\star} - (C+1)}\right)\right)^2 \;. \label{eq:r 1/beta}
\end{equation}
Next, we characterize when $\pi^{\star}(r^\star) > r^\star$, i.e., the customers must pay to adopt the product, instead of the alternative, in which they essentially are paid to adopt.
Let $\hat{r}$ be such that $\hat{r} = \pi^\star(\hat{r})$ and define $x \coloneqq \hat{r} - (C + 1)$. The principal branch of the Lambert W-function satisfies $W(z) = \ln(z) - \ln(W(z))$ for $z > 0$, and hence, from \eqref{eq:pi star firm},
\begin{align*}
    W(e^{p + qF_0 + x}) = x 
    \implies
    e^{p + qF_0} = W(e^{p + qF_0 + x}) = x \;. 
\end{align*}
% where \eqref{eq:lambertw ln} follows from the following property of the principal branch of the Lambert W function when $z > 0$:
% \begin{align*}
%     W(z) = \ln(z) - \ln(W(z)) \;.
% \end{align*}
% Comparing \eqref{eq:compare e^{p + qF}} to \eqref{eq:compare x}, $x = e^{p + qF}$. 
Thus, $\hat{r} = C + 1 + e^{p + qF_0}$.
From \eqref{eq:pi star firm}, $\pi^\star(0) > 0$ and
\begin{align*}
    \frac{\partial \pi^\star(r)}{\partial r} = \frac{W(e^{p + qF_0 + r - (C+1)})}{1 + W(e^{p + qF_0 + r - (C+1)})} \in (0,1) \;. 
\end{align*}
Thus, if $r < \hat{r}$, then $\pi^\star(r) > r$ and similarly, if $r > \hat{r}$, then $\pi^\star(r) < r$. Let $\hat{\beta}$ be such that $r^\star = \hat{r}$. Using \eqref{eq:pi star firm} and the fact that $\pi^\star(\hat{r}) = \hat{r}$ in \eqref{eq:r 1/beta}, 
\begin{gather*}
     \frac{1}{\hat{\beta}} = \hat{r} + \left(\hat{r} - C\right)^2
     = C + 1 + e^{p + qF_0} + \left(1 + e^{p + qF_0}\right)^2 \;.
\end{gather*}
We have already shown that $r^\star$ decreases with $\beta$, so for $\beta > \hat{\beta}$, $r^\star < \hat{r}$, and consequently, $\pi^\star(r^\star) > r^\star$.

Finally, from \eqref{eq:F+ with r}, the optimal final adoption level for the single-period Stackelberg game is 
\begin{equation}
    F_1 = F_0 + (1-F_0)R\left(F_0, \pi^\star(r^\star), r^\star\right) \;. \label{eq:F1}
\end{equation}
The only term in \eqref{eq:F1} which depends on $r^\star$ is the final one. Thus, from \eqref{eq:B and R with r} and \eqref{eq:B pi star with r},
\begin{align*}
R\left(F_0, \pi^\star(r^\star), r^\star\right)
 &= \frac{W\left(e^{p + qF_0 + r^\star - (C+1)}\right)}{1 + W\left(e^{p + qF_0 + r^\star - (C+1)}\right)} \\
\frac{d R\left(F_0, \pi^\star(r^\star), r^\star\right)}{d r^\star} & = \frac{W\left(e^{p + qF_0 + r^\star - (C+1)}\right)}{\left(1 + W\left(e^{p + qF_0 + r^\star - (C+1)}\right)\right)^3} > 0 \;.
\end{align*}
From the last statement, and returning to \eqref{eq:F1}, we see that $F_1$ increases for increasing $r^\star$.
%\Halmos
\endproof

Theorem \ref{thm:game} shows that the optimal rebate, $r^\star$, is unique for the single-stage game, even though it is difficult to find a closed-form solution for it. We find conditions under which the optimal price $\pi^\star$ exceeds $r^\star$ along the equilibrium path for this game, i.e., the customers incur a net cost to adopt. Furthermore, we show that the presence of rebates does in fact increase total adoption. 
% Finally, we perform sensitivity analysis for $r^\star$ with respect to model parameters $\beta$, $p$, and $q$. 

\begin{figure}[!h]
\centering
\includegraphics[scale=0.6]{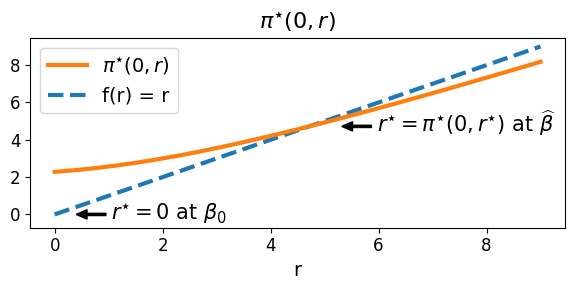}
\caption{Optimal price for the single-period game for parameters $p = q = C = 1, F = 0$. When $\hat{\beta} < \beta < \beta_0$, we have $0 < r^\star < \pi^\star(r^\star)$.}
\label{fig:theorem game}
\end{figure}

Theorem \ref{thm:game} suggests that a policymaker should set $\beta \in \left(\hat{\beta}, \beta_0\right)$ if they seek to provide a rebate which provides some financial assistance to consumers, but does not eliminate their purchase price altogether. Figure \ref{fig:theorem game} illustrates these results.

Recall that $\beta$ measures the relative importance that the policymaker puts on increasing adoption of the product in the population to handing out rebates. 
Theorem \ref{thm:game} tells us that when $\beta$ is small, it is optimal for the policymaker to offer a rebate to spur adoption, but to withhold it when $\beta$ crosses a threshold. In the latter case, the policymaker refuses to influence the market through rebates and simply allows the firm to optimize its prices to maximize its profit.

Next, we analyze the sensitivity of $r^{\star}$ to parameters $\beta$, $p$, and $q$. The right-hand side of \eqref{eq:r 1/beta} increases with $r^{\star}$, so $r^\star$ decreases with increasing $\beta$. This agrees with our intuition that if the policymaker places greater value on a reduction in rebate, the optimal rebate must decrease. 
Again from \eqref{eq:r 1/beta}, if $p$ or $q$ increases while holding $\beta$ constant, $r^\star$ must decrease. An increase in the coefficients of innovation and imitation amounts to an amplification of the tendency of the customers to adopt the product by themselves. Such tendencies offset the need for incentives through rebates for similar adoption levels.

In our model, the customers face a net cost of $\pi^{\star}(r^\star) - r^\star$, where $\pi^{\star}(r^\star)$ is the optimal price for the single-stage game. 
%\bose{Drop $F$ in $\pi^\star(r^\star)$.} 
Theorem \ref{thm:game} provides a lower threshold for $\beta$ below which customers do not pay, but are in fact paid to adopt the product. One presumes such scenarios to rarely occur in practice. We remark that the Illinois Solar for All program, launched in 2019, offering solar installations with no upfront costs to income-eligible households per \citet{illinois_solar_for_all}, stands as an example of that scenario. Our adoption dynamics in \eqref{eq:B and R with r} is such that it varies smoothly with the net cost. Yet when $\pi^{\star}(r^\star) \leq r^\star$, i.e., when the product is free and/or customers are paid to adopt, we may in reality expect a spike in adoption from at least a portion of the population, which may not have adopted otherwise. In essence, we do not expect our model to accurately reflect the adoption dynamics when $\beta \leq \hat{\beta}$.
% , the model's predictive power might decrease substantially. 

From \eqref{eq:beta hat}, it follows that $\hat{\beta}$ decreases with increasing $p$ and $q$. We have already shown in our earlier sensitivity analysis that the optimal rebate decreases with increasing $p$ and $q$. Thus, for a fixed $\beta$, by increasing $p$ or $q$, $r^{\star}$ decreases, whereas from \eqref{eq:pi star firm}, $\pi^{\star}(r)$ increases for any $r$. 
%\bose{Take care of $\pi$} 
Thus, in order to increase the optimal rebate relative to optimal price, $\beta$, the emphasis on reducing the rebate must be lowered. Finally, note from \eqref{eq:beta 0} that $\beta_0$ also decreases with increasing with $p$ and $q$. Thus, by increasing $p$ or $q$, the interval $\left(\hat{\beta}, \beta_0\right)$ shrinks.

% \bose{Predictive power of the model when prices are negative is an interesting point. Leave a comment about the same in the section where the model is being developed and prepare the reader to see more about it here. Also, expand it using perhaps the motivation before Proposition 4. }

\subsection{Numerical Simulations of the $T$-Period Game} \label{sec: numerical game}
While we are not able to analytically characterize similar properties of $r^\star$ as in Theorem \ref{thm:game} for $T > 1$, our numerical simulations in the sequel suggest that the qualitative insights carry over. We used the same methods as in Section \ref{sec:special case} to simulate the optimal trajectories of the firm, and repeated this process for multiple rebate values $r$ in order to find an optimal $r^\star$. 

% \bose{This part is severely underdeveloped. Show things that are similar to the one-stage game (which you have kind of done in the figures, but can be expanded, e.g., show how $\beta$ impacts the adoption paths), and then show some behaviors that are perhaps a bit different from the one-stage game (check things with high $q$'s and see if you can come with some behaviors that are not similar to the $T=1$ case). The second part will take more ingenuity. We have to basically argue whether to expect the same type of results as the one-stage game or not. For example, you can see what happens to the $r^\star$ by sweeping through $\beta$ and check whether there is a lower threshold for $\beta$ for prices to be higher than the rebate. This is just one example. Think more creatively.} 

\begin{figure}[!h]
\centering
\includegraphics[scale=0.58]{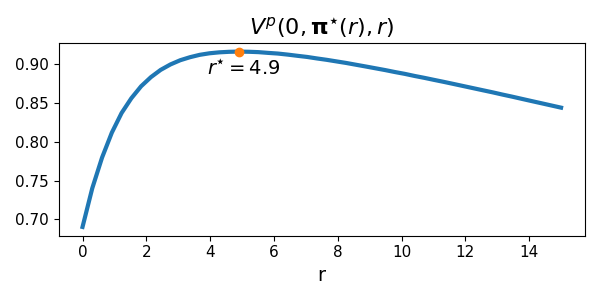}
\caption{Policymaker objective for various rebate values for case $p = q = C = 1, \beta = 0.01, T = 5$. The unique optimal rebate is also marked.}
\label{fig:unique r star}
\end{figure}

\begin{figure}[!h]
\centering
\includegraphics[scale=0.6]{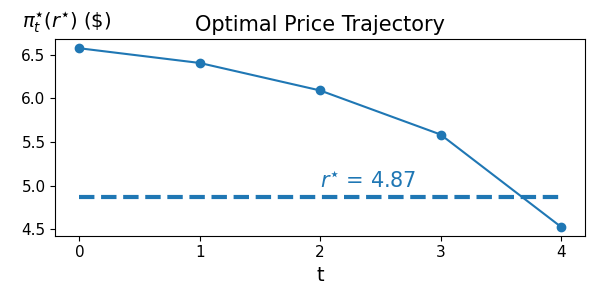}
\caption{Optimal pricing trajectory for parameters $p= q =0.5, C = 0, \beta = 0.01, T = 5$. The dashed line displays the optimal rebate level.}
\label{fig:pi below r star}
\end{figure}

\begin{figure}[!h]
\centering
\includegraphics[scale=0.6]{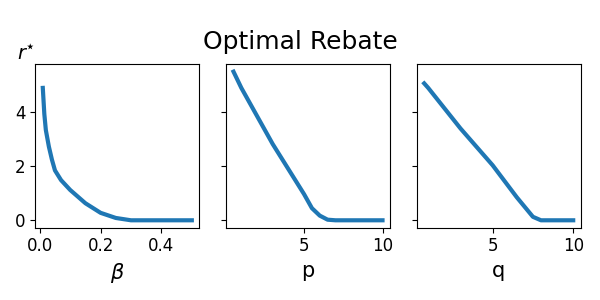}
\caption{Optimal rebate for various values of $\beta$, $p$, and $q$. The common parameters are $p = q = C = 1, \beta = 0.01, T = 5$, with one parameter varying for each scenario.}
\label{fig:beta to r star}
\end{figure}

The optimal rebate is unique for all the cases we simulated, which agrees with our results in Theorem \ref{thm:game} for the single-period game. Figure \ref{fig:unique r star} displays such an example. In addition, for $\beta$ sufficiently small, we encounter the special case of the optimal pricing trajectory dipping below $r^\star$, as shown in Figure \ref{fig:pi below r star}. Furthermore, the sensitivity analysis for $r^\star$ with respect to model parameters following Theorem \ref{thm:game} again appear to apply for the multi-period Stackelberg game. Figure \ref{fig:beta to r star} shows the variation of $r^\star$ over different values of $\beta$, $p$, and $q$; they reflect the trends we deduced from Theorem \ref{thm:game} for the single-period game.  

\begin{figure}[!h]
\centering
\includegraphics[scale=0.6]{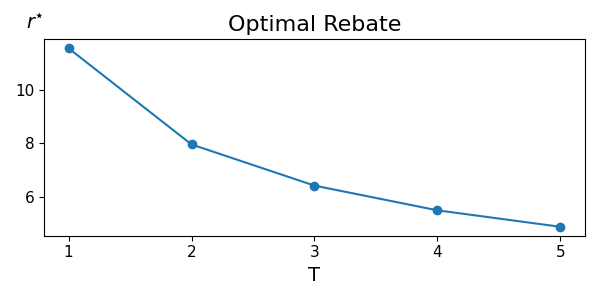}
% \vspace{0.1cm}
\includegraphics[scale=0.6]{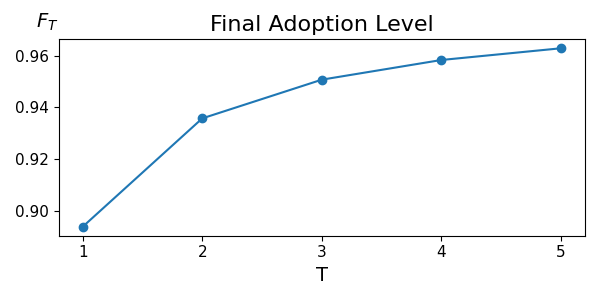}
\caption{Optimal rebate (top) and final adoption level (bottom) for various time horizons $T$. The common parameters are $p = q = C = 1, \beta = 0.01, T = 5, F_0 = 0$.}
\label{fig:over T}
\end{figure}

Our simulations suggest that by increasing the time horizon $T$, $r^\star$ decreases. An explanation for this phenomenon is that since there are more chances to increase adoption in a longer time frame, the policymaker can afford to reduce the rebate level and allow its effects to propagate for some time. On the other hand, in a shorter time period, there are fewer chances to increase adoption, so the policymaker must increase financial incentives to quickly make the product more attractive to customers. Finally, increasing the time horizon results in higher total adoption levels, which agrees with our intuition that over a longer time period, fluctuations in price and population adoption level prompt different subsets of the population to adopt. Figure \ref{fig:over T} displays these results. In essence, all our qualitative insights and policy-relevant discussions from the single-period game carry over to the multi-period setting, showcasing the power of our analysis in Theorem \ref{thm:game} made possible by the nature of our adoption model developed in \eqref{eq:F+ complete}.

\section{Conclusions}
In this paper, we introduced a Bass-inspired adoption model, which we then used to formulate an optimal pricing problem. Our model was shown to be amenable to mathematical analysis and we were able to prove structural results on the optimal pricing policies. Furthermore, it fit well to multiple datasets of product adoption. We then formulated a Stackelberg game between a policymaker and a monopoly using the adoption model to describe consumer response to rebates offered and prices set by the policymaker and the monopolist, respectively. We analytically characterized crucial properties of the optimal rebate in the single-period game, and illustrated that the qualitative properties continue to hold for the multi-stage game.
% Our game formulation provides tool for policymakers and firms to determine optimal rebate and pricing strategies.

There are a number of interesting directions for future research. In our work, we only considered a single firm that sets its price to maximize its profit from the sales of a new product. Marketplaces today seldom sport a single monopolist; even for technology-intensive products, an oligopoly often emerges, e.g., the EV market. Competition among firms and consumer responses to slight variations in the product offerings from said competing firms dynamically interact with rebates and subsidies offered over time. Moreover, rebates are often offered through tax credits or as fixed percentages of purchase prices (as for residential solar). Modeling such nuances in diffusion dynamics is left for future work. Finally, of independent theoretical interest to us is to explore a rigorous framework through which aggregate diffusion models can be derived as aggregations of a large population of rich agent-based models, often used in practice to study product adoption paths.

\appendix
\section{Appendix: Proof of Proposition \ref{prop:general}} \label{proof of prop:general}
Consider the claim $\mathcal{S}_t$: $V_{t+1}$ is 
$L_{V_{t+1}}$-Lipschitz continuous for some $0 < L_{V_{t+1}} < \infty$. We first show that $\mathcal{S}_{T-2}$ holds. Then, we argue that $\mathcal{S}_t$ implies that an optimal price $\pi^\star_t \in [0, \overline{\pi}_t]$ and $V_t(F) \geq V_{t+1}(F)$ for each time $t=0, \ldots, T-2$. Finally, we establish that $\mathcal{S}_{t}$ implies $\mathcal{S}_{t-1}$. The result then follows from induction.

Note that from Theorem \ref{thm:pi* T-1}, $\pi^\star_{T-1}(F)$ is unique and finite.

% We prove the claims via induction. First, we establish the base case for $t = T-1$. For each $F \in [0,1]$, the unique $\pi^{\star}_{T-1}$ is finite and given by \eqref{eq:pi* T-1}. 

$\bullet$ Proving $\mathcal{S}_{T-2}$. 

Note that
\begin{subequations}
\begin{align}
    V_{T-1}(F) 
    % & = H(F, \pi^{\star}_{T-1}(F)) \\
    & = (1-F)\left(\pi^{\star}_{T-1}(F)-C\right)R(F, \pi^{\star}_{T-1}(F)) \\
    & = (1-F) e^{ p + qF - \pi^{\star}_{T-1}(F) }\label{eq:refer G pi *} \\
    & = (1-F)W(e^{p + qF - (C+1)}) \;, \label{eq:J T-1}
\end{align}
\end{subequations}
where \eqref{eq:refer G pi *} follows from \eqref{eq:G pi *} and \eqref{eq:J T-1} from the property $W(z) = ze^{-W(z)}$. Since $W'(z) = W(z)/(z(1+W(z)))$,
\begin{equation}
V_{T-1}'(F) = W(e^{p + qF - (C+1)}) \left(\frac{(1-F)q}{1 + W(e^{p + qF - (C+1)})} - 1\right). \label{eq:dJ T-1}
\end{equation}
Thus, $\forall F \in [0,1]$,
\begin{align*}
    \left|V_{T-1}'(F)\right| \leq W\left(e^{p + q - (C+1)}\right) \left(\frac{q}{1 + W(e^{p - (C+1)})} + 1\right) \;.
\end{align*}
Since $\left|V_{T-1}'(F)\right|$ remains bounded over $[0,1]$, the claim $\mathcal{S}_{T-2}$ holds with $L_{{T-1}} := \max_{F \in [0,1]}\left|V_{T-1}'(F)\right|$.

$\bullet$ Proving $\mathcal{S}_t \implies \pi^\star_t \in [0, \overline{\pi}_t]$ for some $\overline{\pi}_t$.

 Let $F^+(F, \pi)$ govern the dynamics of the proportion of adoption in the population:
\begin{equation}
    F^+(F_t, \pi_t) \coloneqq F_t + R(F_t, \pi_t)(1-F_t) \;. \label{eq:F+} 
\end{equation}
Let $Q_t(F_t, \pi_t)$ be the state-action value function at time $t$,
\begin{equation}
    \begin{gathered}
    Q_{T-1}(F_{T-1}, \pi_{T-1}) = H(F_{T-1}, \pi_{T-1}), \\
    Q_t(F_t, \pi_t) = H(F_t, \pi_t) + V_{t+1}(F^+(F_t, \pi_t))\;,
    \end{gathered}
    \label{eq:Q}
\end{equation}
for $t = 0, \ldots, T-2$  with $H$ defined in \eqref{eq:H}.
We first show that $Q_t$ is sandwiched between two functions which are both monotonically decreasing to 0 for large $\pi$. We then find particular $\underline{\pi_t}$ and $\overline{\pi}_t$ such that $Q_t(F,\underline{\pi_t})$ is at least as large as its upper bound for $\pi \geq \overline{\pi}_t$, thus allowing us to only consider $\pi \leq \overline{\pi}_t$ when maximizing $Q_t(F,\pi)$.

%\bose{Define the notation $F^+$. Look for my comment around (31).}
Assume $\mathcal{S}_t$, $V_{t+1}(F)$ is $L_{V_{t+1}}$-Lipschitz continuous over $[0,1]$. Thus, 
    \begin{equation}
    \left|V_{t+1}(F^+(F,\pi)) - V_{t+1}(F)\right| \leq L_{V_{t+1}}\left|F^+(F,\pi) - F\right| = L_{V_{t+1}}(1-F)R(F, \pi) .  \label{eq:L_{V_{t+1}}}
    \end{equation}
    Then, from \eqref{eq:Q} and \eqref{eq:L_{V_{t+1}}},
    \begin{equation}
    \begin{aligned}
        Q_t(F,\pi) & \geq H(F,\pi) + V_{t+1}(F) - L_{V_{t+1}}(1-F)R(F, \pi) \\
        & = \underline{H}(F,\pi) + V_{t+1}(F), \qquad \forall F \in [0,1]\;,
    \end{aligned} \label{eq:Q lower bound}
    \end{equation}
    where $\underline{H}(F,\pi) = (1-F)R(F, \pi)\left[\pi-(C + L_{V_{t+1}})\right]$. 
    Similarly,
    \begin{equation}
        \begin{aligned}
        Q_t(F,\pi) \leq \overline{H}(F,\pi) + V_{t+1}(F), \quad \forall F \in [0,1] \;,
    \end{aligned}\label{eq:Q upper bound}
    \end{equation}
    where $\overline{H}(F,\pi) = (1-F)R(F, \pi)\left[\pi-(C - L_{V_{t+1}})\right]$.
    Note that $\underline{H}$ and $\overline{H}$ both equal $H$ with cost $C$ replaced by $C + L_{V_{t+1}}$ and $C - L_{V_{t+1}}$, respectively. 
Let $\underline{\pi_t} = C + L_{V_{t+1}} + 1$. Then, from \eqref{eq:Q lower bound}, $\forall F \in [0,1]$,
\begin{equation}
\begin{aligned}
    Q_t(F,\underline{\pi_t}) & \geq (1-F)R(F, \underline{\pi_t}) + V_{t+1}(F) \\
    & \geq (1-F)\underbrace{R(0, \underline{\pi_t})}_{>0} + V_{t+1}(F) \;,
\end{aligned} \label{eq:R(0,pi)}
\end{equation}
where the last line is due to ${\partial R(F, \pi)}/{\partial F} \geq 0 $.
% $\frac{\partial R(F, \pi)}{\partial F} = \frac{R(F, \pi)}{1 + B(F, \pi)}q \geq 0 $.
Using an argument similar to that in the proof of Theorem \ref{thm:pi* T-1}, for any fixed $F \in [0,1)$, $\overline{H}(F,\pi)$ attains its maximum at 
    \begin{equation}
        \overline{\pi}^{\star}(F) = C - L_{V_{t+1}} + 1 + W\left(e^{p + qF - (C- L_{V_{t+1}}+1)}\right) \;, 
    \label{eq:pi overbar*(F)}
    \end{equation}
    after which it decreases monotonically to $0$. Because $\partial R(F, \pi)/\partial F \geq 0$, the following holds for all $F \in [0,1]$:
    \begin{equation}
        \overline{H}(F,\pi) \leq (1-F)R(1, \pi)\left[\pi-(C - L_{V_{t+1}})\right] \;. \label{eq:R(1,pi)}
    \end{equation}
From the analysis of $H$ in the proof of Theorem \ref{thm:pi* T-1}, $R(1, \pi)\left[\pi-(C - L_{V_{t+1}})\right]$ decreases monotonically to $0$ over $\pi$ for $\pi > \overline{\pi}^{\star}(1)$. Furthermore, because $R(0,\underline{\pi_t}) > 0$, the following is finite:
\begin{equation}
\overline{\pi}_t = \min\left\{\pi \geq \max\left\{\underline{\pi_t}, \overline{\pi}^{\star}(1)\right\}\right. ~\big|~ \left.R(1, \pi)\left[\pi-(C - L_{V_{t+1}})\right] \leq R(0, \underline{\pi_t}) \right\} \;. \label{eq:pi overbar t}
\end{equation}
    Note that from \eqref{eq:pi overbar*(F)}, $\overline{\pi}^{\star}(F)$ attains its maximum over $[0,1]$ at $F = 1$, so $\overline{\pi}_t \geq \overline{\pi}^{\star}(F), \forall F \in [0,1]$. Because $R(1, \pi)\left[\pi-(C - L_{V_{t+1}})\right]$ decreases monotonically to $0$ for $\pi > \overline{\pi}_t \geq \overline{\pi}^{\star}(1)$,
    \begin{equation}
        R(1, \pi)\left[\pi-(C - L_{V_{t+1}})\right] \leq R(0, \underline{\pi_t}), \quad \forall \pi \geq \overline{\pi}_t \;. \label{eq:R(1, pi) R(0,pi)}
    \end{equation}
    Then, for all $F \in [0,1]$ and $\pi \geq \overline{\pi}_t$,
    \begin{subequations}
    \begin{align}
        Q_t(F,\pi) & \leq \overline{H}(F,\pi) + V_{t+1}(F) \label{proof:refer Q upper bound} \\
    & \leq (1-F)R(1, \pi)\left[\pi-(C - L_{V_{t+1}})\right] + V_{t+1}(F) \label{proof:refer R(1,pi)}\\
    & \leq (1-F)R(0, \underline{\pi_t}) + V_{t+1}(F) \label{proof:refer R(1, pi) R(0,pi)}\\
    & \leq Q_t(F,\underline{\pi_t})  \;, \label{proof:refer R(0,pi)}
    \end{align}
    \end{subequations}
    where \eqref{proof:refer Q upper bound} follows from \eqref{eq:Q upper bound}, \eqref{proof:refer R(1,pi)} from \eqref{eq:R(1,pi)}, \eqref{proof:refer R(1, pi) R(0,pi)} from \eqref{eq:R(1, pi) R(0,pi)}, and \eqref{proof:refer R(0,pi)} from \eqref{eq:R(0,pi)}. Thus, the search for the global optimizer(s) of $Q_t(F,\pi)$ can be restricted to be over $\pi \in [0,\overline{\pi}_t]$:
    \begin{align*}
        \sup_{\pi \geq 0} Q_t(F, \pi) \geq Q_t(F,\underline{\pi_t}) \geq Q_t(F,\pi), \quad \forall \pi \geq \overline{\pi}_t \;.
    \end{align*}
    By the induction hypothesis, $V_{t+1}$ is Lipschitz continuous, so $Q_t(F,\pi)$, a composition of continuous functions, is also continuous, and hence, 
    % . Since a continuous function achieves its maximum over a closed interval, 
    \begin{align*}
        V_t(F) 
        % = \sup_{\pi \geq 0} Q_t(F, \pi) 
        = \max_{\pi \in [0, \overline{\pi}_t]} Q_t(F,\pi) \;.
    \end{align*}
    
$\bullet$ Proving $\mathcal{S}_t \implies V_t(F) \geq V_{t+1}(F)$. 

From the result of the previous step, 
    \begin{align*}
        V_t(F) = \max_{\pi \in [0, \overline{\pi}_t]} Q_t(F,\pi) \geq Q_t(F, C + L_{V_{t+1}}) \geq V_{t+1}(F) \;,
    \end{align*}
    where the first inequality holds because $C + L_{V_{t+1}} < C + L_{V_{t+1}} + 1 = \underline{\pi_t} \leq \overline{\pi}_t$ (see \eqref{eq:pi overbar t}), and the last inequality follows from \eqref{eq:Q lower bound}. 
    As a direct consequence,
    \begin{align*}
     Q_t(F, \pi) \geq H(F, \pi) + V_{t + 2}(F^+(F,\pi)) = Q_{t+1}(F, \pi) \;.
    \end{align*}
    
$\bullet$ Proving $\mathcal{S}_t \implies \mathcal{S}_{t-1}$.

    We first show that $H(F,\pi)$ is Lipschitz continuous. Note that
\begin{align}
    \frac{\partial H(F,\pi)}{\partial F} = \left(\pi-C\right)R(F, \pi)\left[\frac{(1-F)q}{1 + B(F, \pi)} - 1\right].
    \label{eq:dH.dF}
\end{align}
From the analysis of $H$ in the proof of Theorem \ref{thm:pi* T-1}, $\left(\pi-C\right)R(F, \pi)$ strictly increases then decreases as a function of $\pi$ over $\mathbb{R}^+$, with a limit of $0$ as $\pi \rightarrow \infty$. Furthermore, $\left(\pi-C\right)R(F, \pi)$ is finite at $\pi = 0$, so $\left(\pi-C\right)R(F, \pi)$ is bounded over $\pi \in \mathbb{R}^+$ and $F \in [0,1]$. Since $(1-F)/\left(1 + B(F, \pi)\right) \in [0,1]$ over $\pi \in \mathbb{R}^+$ and $F \in [0,1]$ and $q$ is finite, $\left|\partial H(F,\pi)/\partial F\right|$ is bounded. Let $L_H = \sup_{\pi \in \mathbb{R}^+, F \in [0,1]}\left|\partial H(F,\pi)/\partial F\right| \in \mathbb{R}^+$ so that $H(F,\pi)$ is $L_H$-Lipschitz in $F$, uniformly over $\pi \in \mathbb{R}^+$. For all $\tilde{F},F \in [0,1]$, 
\begin{equation}
    \left|H(\tilde{F},\pi) - H(F,\pi)\right| \leq L_H|\tilde{F}-F|\;. \label{eq:LH}
\end{equation}
%\bose{Take this up I suppose!} 
Note that
\begin{equation}
\frac{\partial F^+(F, \pi)}{\partial F} = 1 - \underbrace{R(F, \pi)}_{\in[0,1]} + \underbrace{(1-F)\frac{R(F, \pi)}{1 + B(F, \pi)}}_{\in[0,1]}q \geq 0 \;, \label{eq:dF+dF}
\end{equation}
which is also bounded over $\pi \in \mathbb{R}^+$ and $F \in [0,1]$. Letting $L_{F^+} = \sup_{\pi \in \mathbb{R}^+, F \in [0,1]}\left|\partial F^+(F, \pi)/\partial F\right| \in \mathbb{R}^+$, $F^+(F, \pi)$ is $L_{F^+}$-Lipschitz in $F$, uniformly over $\pi \in \mathbb{R}^+$. For all $\tilde{F},F \in [0,1]$,
\begin{equation}
    \left|F^+(\tilde{F},\pi) - F^+(F,\pi)\right| \leq L_{F^+}|\tilde{F}-F| \;. \label{eq:LF+}
\end{equation}
    
Using the induction hypothesis and \eqref{eq:LF+}, we have that $\left|V_{t+1}(F^+(\tilde{F}, \pi)) - V_{t+1}(F^+(F, \pi))\right| \leq L_{V_{t+1}}L_{F^+}|\tilde{F}-F|, \forall \tilde{F},F \in [0,1]$.
    Using \eqref{eq:LH} and letting $L_{V_t} = L_H + L_{V_{t+1}}L_{F^+} \in \mathbb{R}^+$,
    \begin{equation}
    \begin{aligned}
        \left|Q_t(\tilde{F},\pi) - Q_t(F, \pi)\right|
        & \leq \left|H(\tilde{F},\pi) - H(F,\pi)\right| + \left|V_{t+1}(F^+(\tilde{F}, \pi)) - V_{t+1}(F^+(F, \pi))\right| \\
        & \leq L_{V_t}|\tilde{F}-F|, \qquad\forall \tilde{F},F \in [0,1] \;.
    \end{aligned}
    \label{eq:Lipschitz Q}
    \end{equation} 
    We have already shown that $V_t(F) = \max_{\pi \in [0, \overline{\pi}_t]} Q_t(F,\pi), \forall F \in [0,1]$. For any $F, \tilde{F} \in [0,1]$, let $\tilde{\pi}^{\star} \in \arg\max_{\pi \in [0, \overline{\pi}_t]} Q_t(\tilde{F},\pi)$. Then,
    \begin{subequations}
    \begin{align}
        V_t(F) 
        % & = \max_{\pi \in [0, \overline{\pi}_t]} Q_t(F,\pi) \\
        & \geq Q_t(F,\tilde{\pi}^{\star})\\
        & \geq Q_t(\tilde{F},\tilde{\pi}^{\star}) - L_{V_t}|\tilde{F} - F| \label{eq:refer Lipschitz Q} \\
        & = V_t(\tilde{F}) - L_{V_t}|\tilde{F} - F| \;,
    \end{align}   
    \end{subequations}
    where \eqref{eq:refer Lipschitz Q} follows from \eqref{eq:Lipschitz Q}. Reversing the roles of $F$ and $\tilde{F}$, we obtain $V_t(\tilde{F}) \geq V_t(F) - L_{V_t}|\tilde{F} - F|$. Thus, $\left|V_t(\tilde{F}) - V_t(F)\right| \leq L_{V_t}|\tilde{F}-F|, \forall \tilde{F},F \in [0,1]$. Hence, $V_t$ is $L_{V_t}$-Lipschitz over $[0,1]$. 
%\Halmos
\endproof

\section{Appendix: Proof of Theorem \ref{thm:special case}} \label{proof of thm:special case}
Throughout this proof, we will use the  definitions of $B$ and $R$ in \eqref{eq:B and R}, $H$ in \eqref{eq:H} and $Q_t$ in \eqref{eq:Q}. Define 
$\mathcal{O} = (-\varepsilon, 1)$ for arbitrarily small $\varepsilon > 0$.
%\bose{Make this equation inline and remove any references to it.} 
We will use an induction argument to prove the claim $\mathcal{S}_t$ at time $t$, which is as follows: 1) $V_t(F) \in \mathcal{C}^{t + 2}$ over $\mathcal{O}$\footnote{Technically, we optimize only over non-negative prices in \eqref{eq:dp}. Conceptually, nothing changes upon allowing the domain of prices to contain $0^{-}$.}, where $\mathcal{C}^k$ is the set of functions whose derivatives up to and including the $k^{\text{th}}$ exist and are continuous, and 2) $V_t'(F) \leq 0$ and $V_t(F)$ is concave over $\mathcal{O}$.

We first show that $\mathcal{S}_{T-1}$ holds. Then, we argue that for $t=0, \ldots, T-2$, $\mathcal{S}_{t+1}$ implies that the optimal price at time $t$, $\pi^{\star}_t(F)$, is unique and positive, $\pi^{\star}_t(F) \in \mathcal{C}^{t + 2}$ over $\mathcal{O}$ and $[\pi^\star_t]'(F) \geq 0$. We subsequently establish that $\mathcal{S}_{t+1}$ implies $\mathcal{S}_t$, and the result follows from induction. Finally, we show as a consequence of the aforementioned results that $\pi^{\star}_t(F) \geq \pi^{\star}_{t+1}(F)$ and $V_t'(F) \leq V_{t+1}'(F), \forall t = 0,\ldots, T-2$.

We showed in Theorem \ref{thm:pi* T-1} that $\pi^{\star}_{T-1}(F) > 0$ and is unique. From \eqref{eq:pi* T-1}, $\pi_{T-1}^{\star}(F)$ is smooth, i.e., infinitely differentiable, so
\begin{align}
    \left[\pi_{T-1}^{\star}\right]'(F) = \frac{W(e^{p + qF - (C+1)})q}{1 + W(e^{p + qF - (C+1)})} \geq 0 \;.
\end{align}
% \bose{Here, include the list of bullet points as the steps required to prove the result.}

We will prove the following results in order to show that $\mathcal{S}_t$ holds. We will begin with the base case and show $\mathcal{S}_{T-1}$. Then, assuming $\mathcal{S}_{t+1}$ to be true, we show the following results in order: 
\begin{enumerate}[label=(\roman*)]
    \item  $\pi^{\star}_t(F)$ is unique and positive.
    \item $ \pi^{\star}_t(F) \in \mathcal{C}^{t + 2}$ over $\mathcal{O}$.
    \item $\left[\pi^{\star}_t\right]'(F) \geq 0$.
    \item $ V_t(F) \in \mathcal{C}^{t + 2}$ over $\mathcal{O}$.
    \item $V_t'(F) \leq 0$ over $\mathcal{O}$.
    \item  $V_t(F)$ is concave over $\mathcal{O}$.
\end{enumerate}

$\bullet$ \textit{Proving $\mathcal{S}_{T-1}$:}

From \eqref{eq:J T-1}, $V_{T-1}(F)$ is smooth. From \eqref{eq:dJ T-1}, $q\leq 1$ implies $V_{T-1}'(F) \leq 0$. Furthermore,
\begin{equation}
V_{T-1}''(F) = \frac{W(e^{p + qF - (C+1)})q}{\left(1 + W(e^{p + qF - (C+1)})\right)^3} \left[(1-F)q - 2\left(1 + W\left(e^{p + qF - (C+1)}\right)\right)^2\right] \;.   \label{eq:d2J T-1}
\end{equation}
From \eqref{eq:d2J T-1}, $q\leq 1$ implies $V_{T-1}(F)$ is concave over $\mathcal{O}$. 

$\bullet$ \textit{Proving $\mathcal{S}_{t+1} \implies$ $\pi^{\star}_t(F)$ is unique and positive:}

At $F = 1$, $Q_t(F, \pi) = 0$ from \eqref{eq:Q}; i.e., there can be no incremental profit since the entire population has adopted. We will focus on the case $F < 1$ in our analysis.

By $\mathcal{S}_{t+1}$, $V_{t+1}(F) \in \mathcal{C}^{t + 3}$ over $\mathcal{O}$. Since $F^+(F, \pi) \in [0,1]$ for $F \in [0,1]$ (see \eqref{eq:F+}) and varies smoothly in both parameters, $Q_t(F, \pi)$ is differentiable in $\pi$ with
\begin{equation}
\frac{\partial Q_t(F, \pi)}{\partial \pi} = (1-F)R(F, \pi)  \left[1 - \underbrace{\frac{\pi-C + V_{t+1}'(F^+(F, \pi))}{1 + B(F, \pi)}}_{\eqqcolon G_Q(\pi)}\right] \;. \label{eq:dQ}
\end{equation}

By $\mathcal{S}_{t+1}$, $V_{t+1}$ is concave over $\mathcal{O}$, i.e. $V_{t+1}'$ is non-increasing. Additionally,
\begin{equation}
    \frac{\partial F^+(F, \pi)}{\partial \pi} = -(1-F)\frac{R(F, \pi)}{1 + B(F, \pi)} \leq 0 \;, \label{eq:dF+dpi}
\end{equation}
so $V_{t+1}'(F^+(F, \pi))$ is non-decreasing in $\pi$. Therefore, the numerator of $G_Q(\pi)$ in \eqref{eq:dQ} monotonically grows unbounded in $\pi \in \mathbb{R}_+$. Furthermore, the denominator of $G_Q(\pi)$ decreases monotonically in $\pi$, being bounded below by unity. By $\mathcal{S}_{t+1}$, 
$V_{t+1}'(F) \leq 0$, and so $G_Q(0) \leq 0$, implying that $G_Q$ crosses unity exactly once from below. Thus, from \eqref{eq:dQ}, $Q_t(F, \pi)$ monotonically increases, reaches a unique maximum, and then decreases in $\pi \in \mathbb{R}_+$, implying that $\pi_t^{\star}(F)$ is unique and positive for $F< 1$ as required.

$\bullet$ \textit{Proving $\mathcal{S}_{t+1} \implies \pi^{\star}_t(F) \in \mathcal{C}^{t + 2}$ over $\mathcal{O}$:}

    From \eqref{eq:dQ}, $\pi^{\star}_t(F)$ is uniquely and implicitly determined by $f_t(F, \pi^{\star}_t(F)) = 0$, where
    \begin{equation}
    f_t(F, \pi) \coloneqq  C + 1 - \pi + B(F, \pi) - V_{t+1}'(F^+(F, \pi)).
    \label{eq:implicit}
    \end{equation}
    By $\mathcal{S}_{t+1}$, $V_{t+1} \in \mathcal{C}^{t + 3}$ over $\mathcal{O}$. Thus, by \eqref{eq:implicit}, $f_t(F, \pi) \in \mathcal{C}^{t + 2}$ over $\mathcal{O} \times \mathbb{R}$. Also by $\mathcal{S}_{t+1}$, $V_{t+1}(F)$ is concave over $\mathcal{O}$, so for all $F \in \mathcal{O}, \pi \in \mathbb{R}$,
    \begin{equation}
    \frac{\partial f_t(F, \pi)}{\partial \pi} =  -1 - \underbrace{B(F, \pi)}_{\geq 0} + \underbrace{V_{t+1}''(F^+(F, \pi))}_{\leq 0}\underbrace{(1-F)\frac{R(F, \pi)}{1 + B(F, \pi)}}_{\geq 0} < 0\;.
    \label{eq:df dpi}
    \end{equation}
    By the implicit function theorem (Section 10.3 of \citet{ift}), $\pi^{\star}_t(F) \in \mathcal{C}^{t + 2}$ is uniquely defined over $\mathcal{O}$.

    $\bullet$ \textit{Proving $\mathcal{S}_{t+1} \implies \left[\pi^{\star}_t\right]'(F) \geq 0$:}
    
    From \eqref{eq:implicit},
    \begin{equation}
    \frac{\partial f_t(F, \pi)}{\partial F} = \underbrace{B(F, \pi)q}_{\geq 0} - \underbrace{V_{t+1}''(F^+(F, \pi)}_{\leq 0}\underbrace{\frac{\partial F^+(F, \pi)}{\partial F}}_{\geq 0} \geq 0\;, \label{eq:df dF}
    \end{equation}
    which follows from \eqref{eq:dF+dF} and the fact that $\mathcal{S}_{t+1}$ implies that $V_{t+1}(F)$ is concave.
    Under the same conditions as in the proof that $\pi^{\star}_t(F) \in \mathcal{C}^{t + 2}$ over $\mathcal{O}$, the implicit function theorem gives $\left[\pi^{\star}_t\right]'(F) = -\frac{\partial f_t(F, \pi^{\star}_t(F))}{\partial F}\big/\frac{\partial f_t(F, \pi^{\star}_t(F))}{\partial \pi}$. By \eqref{eq:df dpi}, $\frac{\partial f_t(F, \pi^{\star}_t(F))}{\partial \pi} \leq -1$. Thus, combined with \eqref{eq:df dF}, $\left[\pi^{\star}_t\right]'(F) \geq 0$.
   
    $\bullet$ \textit{Proving $\mathcal{S}_{t+1} \implies V_t(F) \in \mathcal{C}^{t + 2}$ over $\mathcal{O}$:}
    
    By $\mathcal{S}_{t+1}$, $V_{t+1}(F) \in \mathcal{C}^{t + 3}$ over $\mathcal{O}$, and we already showed that $\pi^{\star}_t(F) \in \mathcal{C}^{t + 2}$ over $\mathcal{O}$. Note that  $F^+(F, \pi) \in \mathcal{O}$ and is smooth for all $(F, \pi) \in \mathcal{O} \times \mathbb{R}$. Furthermore, $H(F, \pi)$ is smooth over $\mathbb{R}^2$, so by composition of $\mathcal{C}^{t + 2}$ functions,
    \begin{align*}
    V_t(F) = H(F, \pi^{\star}_t(F)) + V_{t+1}(F^+(F, \pi^{\star}_t(F))) \in \mathcal{C}^{t + 2}
    \end{align*}
    over $\mathcal{O}$.
    
    $\bullet$ \textit{Proving $\mathcal{S}_{t+1} \implies V_t'(F) \leq 0$ over $\mathcal{O}$:}
    
From \eqref{eq:Q},
\begin{equation}
\begin{aligned}
    V_t'(F) 
    &= \frac{d}{dF}Q_t(F, \pi^{\star}_t(F))
    \\
    &= \frac{\partial}{\partial F}Q_t(F, \pi^{\star}_t(F)) + \cancelto{0}{\frac{\partial}{\partial \pi}Q_t(F, \pi^{\star}_t(F))} [\pi_t^{\star}]'(F)
    \\
     &= \left(\pi^{\star}_t(F)-C\right) R(F, \pi_t^{\star}(F)) \left[ \frac{(1-F)q}{1+B(F, \pi_t^\star(F))} -1 \right] 
     \\
     &\qquad + V_{t+1}'(F^+(F, \pi^{\star}_t(F)))\frac{\partial F^+(F, \pi^{\star}_t(F))}{\partial F}.
\end{aligned}
\label{eq:Jt.prime.1}
\end{equation}
To further simplify the above relation, notice that 
\begin{equation}
\begin{aligned}
    \pi^{\star}_t(F) - C = 1 + B(F, \pi^{\star}_t(F)) - V_{t+1}'(F^+(F, \pi^{\star}_t(F))) 
\end{aligned}\label{eq:pi star}
\end{equation}
from \eqref{eq:implicit} and 
\begin{align*}
    \frac{\partial F^+(F, \pi^{\star}_t(F))}{\partial F} = 1 + R(F, \pi_t^{\star}(F)) \left[ \frac{(1-F)q}{1+B(F, \pi_t^\star(F))} -1 \right] 
\end{align*}
from \eqref{eq:dF+dF}. Using the above two relations in \eqref{eq:Jt.prime.1}, we get
    \begin{equation}
    V_t'(F)= \underbrace{B(F, \pi^{\star}_t(F))}_{\geq 0}\left[\frac{(1-F)q}{1 + B(F, \pi^{\star}_t(F))} - 1\right] + V_{t+1}'(F^+(F, \pi^{\star}_t(F))) \;.
        \label{eq:dJt}
    \end{equation}
    For $\varepsilon$ chosen sufficently small in the definition of $\mathcal{O}$, the result follows from $q \leq 1$ and $V_{t+1}'(F) \leq 0$ (by induction hypothesis). 

    $\bullet$ \textit{Proving $\mathcal{S}_{t+1} \implies V_t(F)$ is concave over $\mathcal{O}$:}
    
    From \eqref{eq:pi star} and \eqref{eq:B}, we have
\begin{equation}
\begin{aligned}
\left[\pi_t^{\star}\right]'(F) = \frac{d}{dF}B(F, \pi^{\star}_t(F)) - \frac{d}{dF}V_{t+1}'(F^+(F, \pi^{\star}_t(F)))
\end{aligned} \label{eq:dpi star}
\end{equation}
\begin{equation}
    \frac{d}{dF}B(F, \pi^{\star}_t(F)) = B(F, \pi^{\star}_t(F))(q - [\pi_t^{\star}]'(F)) \;.
 \label{eq:dB}
\end{equation}
From \eqref{eq:dJt}, 
\begin{subequations}
\begin{align}
    V_t''(F) &= \frac{d}{dF}B(F, \pi^{\star}_t(F))\frac{(1-F)q}{1 + B(F, \pi^{\star}_t(F))} + B(F, \pi^{\star}_t(F))q \frac{d}{dF}\left(\frac{1-F}{1 + B(F, \pi^{\star}_t(F))}\right) \notag
\\
& \qquad -\left(\frac{d}{dF}B(F, \pi^{\star}_t(F)) - \frac{d}{dF}V_{t+1}'(F^+(F, \pi^{\star}_t(F)))\right) \label{eq:sub dpi star}
\\
 &= \frac{d}{dF}B(F, \pi^{\star}_t(F))\frac{(1-F)q}{1 + B(F, \pi^{\star}_t(F))} \left(1 - R\left(F, \pi^{\star}_t(F)\right)\right) - R\left(F, \pi^{\star}_t(F)\right)q - \left[\pi_t^{\star}\right]'(F) 
\label{eq:sub dB}
\\
&= \underbrace{R\left(F, \pi^{\star}_t(F)\right)q}_{\geq 0}\left[\frac{(1-F)q}{1 + B(F, \pi^{\star}_t(F))} -1 - \underbrace{(1-F)[\pi_t^{\star}]'(F)\left(1 - R\left(F, \pi^{\star}_t(F)\right)\right)}_{\geq 0}\right] \notag \\
    &\qquad - \underbrace{[\pi_t^{\star}]'(F)}_{\geq 0} \;,
\end{align}
\end{subequations}
in which we substitute \eqref{eq:dpi star} into \eqref{eq:sub dpi star} and \eqref{eq:dB} into \eqref{eq:sub dB}. The last step shows that $q \leq 1$ is a sufficient condition for $V_t''(F) \leq 0$, implying that $V_t(F)$ is concave over $\mathcal{O}$, for $\varepsilon$ chosen arbitrarily small in the definition of $\mathcal{O}$.
    
    $\bullet$ \textit{Proving $V_t'(F) \leq V_{t+1}'(F)$ for $t = 0, \ldots, T - 2$:}
    
    We have already shown that when $q \leq 1$, $V_t(F)$ is concave over $[0,1]$, and hence, $V_t'(F)$ is non-increasing over $F$, $\forall t = 0, \ldots, T-1$. Thus, for arbitrary $t$,
    \begin{align*}
    V_{t+1}'(F^+(F, \pi^{\star}_t(F))) = V_{t+1}'\left(F + \underbrace{(1-F)R\left(F, \pi^{\star}_t(F)\right)}_{\geq 0}\right)\leq V_{t+1}'(F) \;.   
    \end{align*}
    From \eqref{eq:dJt}, we get the required result as follows:
    \begin{align*}
    V_t'(F) \leq \underbrace{B(F, \pi^{\star}_t(F))}_{\geq 0}\underbrace{\left[\frac{(1-F)q}{1 + B(F, \pi^{\star}_t(F))} - 1\right]}_{\leq 0} + V_{t+1}'(F) \leq V_{t+1}'(F)
    \;.
    \end{align*}
    
$\bullet$ \textit{Proving $\pi^{\star}_t(F) \geq \pi^{\star}_{t+1}(F), \forall t = 0, \ldots, T - 2$:}
    
    For some $F \in [0,1)$, assume to the contrary that $\pi^{\star}_t(F) < \pi^{\star}_{t+1}(F)$.
    From \eqref{eq:implicit} and the last claim,  we have
    \begin{align}
    \begin{aligned}
        \pi^{\star}_{t+1}(F) 
        &= C + 1 + B(F, \pi^{\star}_{t+1}(F)) - V_{t+2}'(F^+(F, \pi^{\star}_{t+1}(F)))
        \\
        &\leq C + 1 + B(F, \pi^{\star}_{t+1}(F))  - V_{t+1}'(F^+(F, \pi^{\star}_{t+1}(F)))
        \\
        &\leq C + 1 + B(F, \pi^{\star}_t(F)) - V_{t+1}'(F^+(F, \pi^{\star}_t(F))) 
        \\
        &= \pi^{\star}_t(F).
    \end{aligned}
    \end{align}
        Here, the second-to-last line follows from the concavity of $V_{t+1}$ (implying $-V_{t+1}'$ is non-decreasing), $\partial F^+(F, \pi)/\partial \pi \leq 0$ from \eqref{eq:dF+dpi}, 
        $\partial B(F, \pi)/\partial \pi \leq 0$ from direct calculation, and our hypothesis $\pi^{\star}_t(F) < \pi^{\star}_{t+1}(F)$. We arrive at a contradiction. 
%\Halmos
\endproof

%----------------------
\bibliographystyle{informs2014}
\bibliography{refs} 

@article{mitra2019forecasting,
  title={Forecasting the Diffusion of Innovative Products Using the Bass Model at the Takeoff Stage: A Review of Literature from Subsistence Markets.},
  author={Mitra, Suddhachit},
  journal={Asian Journal of Innovation \& Policy},
  volume={8},
  number={1},
  year={2019}
}

@article{yu2016market,
  title={Market dynamics and indirect network effects in electric vehicle diffusion},
  author={Yu, Zhe and Li, Shanjun and Tong, Lang},
  journal={Transportation Research Part D: Transport and Environment},
  volume={47},
  pages={336--356},
  year={2016},
  publisher={Elsevier}
}

@article{zapata2024diffusion,
  author    = {Sebastian Zapata and Duvan A. Gomez and Andres Julian Aristizabal and Monica Castaneda and Jorge I. Romero-Gelves},
  title     = {Assessing the diffusion of photovoltaic technology and electric vehicles using system dynamics modeling},
  journal   = {Energy Exploration \& Exploitation},
  pages     = {1--19},
  year      = {2024},
  publisher = {SAGE Publications},
  doi*      = {10.1177/01445987241277917},
  url*      = {https://journals.sagepub.com/doi/pdf/10.1177/01445987241277917}
}

@inproceedings{wang2024diffusion,
  author    = {Xiuchun Wang and Xuedong He and Xiaoqian Sun and Meicui Qin and Ruiping Pan and Yuanyuan Yang},
  title     = {The Diffusion Path of Distributed Photovoltaic Power Generation Technology driven by Individual Behavior},
  booktitle = {Proceedings of the 4th International Conference on Economic Management and Big Data Applications (ICEMBDA 2023)},
  year      = {2024},
  publisher = {EAI},
  address   = {Tianjin, China},
  doi*       = {10.4108/eai.27-10-2023.2341959},
  url*       = {https://eudl.eu/doi/10.4108/eai.27-10-2023.2341959}
}

@article{pathania2017solar,
  author    = {Ajay Kumar Pathania and Bindiya Goyal and Jagdish Raj Saini},
  title     = {Diffusion of adoption of solar energy – a structural model analysis},
  journal   = {Smart and Sustainable Built Environment},
  volume    = {6},
  number    = {2},
  pages     = {66--83},
  year      = {2017},
  publisher = {Emerald Publishing Limited},
  doi*       = {10.1108/SASBE-11-2016-0033},
  url*       = {https://www.emerald.com/insight/content/doi/10.1108/sasbe-11-2016-0033/full/html}
}

@article{xia2022evadoption,
  author    = {Zhengwei Xia and Dongming Wu and Langlang Zhang},
  title     = {Economic, Functional, and Social Factors Influencing Electric Vehicles’ Adoption: An Empirical Study Based on the Diffusion of Innovation Theory},
  journal   = {Sustainability},
  volume    = {14},
  number    = {10},
  pages     = {6283},
  year      = {2022},
  publisher = {MDPI},
  doi*       = {10.3390/su14106283},
  url*       = {https://www.mdpi.com/2071-1050/14/10/6283}
}

@misc{electrek2025stateev,
  author = {Doll, Scooter},
  title = {Every electric vehicle tax credit rebate available, by state},
  year = {2025},
  url = {https://electrek.co/ev-tax-credit-rebate-states-electric-vehicles/},
  note = {Accessed October 21, 2025}
}

@misc{ca_investment_tax_credit,
author = {{Canada Revenue Agency}},
  title = {Clean Technology (CT) Investment Tax Credit (ITC)
    },
  year = {2024},
  url = {https://www.canada.ca/en/revenue-agency/services/tax/businesses/topics/corporations/business-tax-credits/clean-economy-itc/clean-technology-itc/about-ct-itc.html},
  note = {Accessed October 21, 2025}
}

@misc{vecharged2025europeev,
  author = {Suhas},
  title = {Europe’s New 2025 EV Incentives: How to Get Thousands in Savings},
  year = {2025},
  url = {https://vecharged.com/news/europe-electric-car-incentives-2025/},
  note = {Accessed October 21, 2025}
}

@misc{ev_by_country,
  author = {Nikhil Tiwari},
  title = {Electric Vehicle Incentives by Country: Check the Incentives of USA, European Union, China, Indian, Canada, and United Kingdom},
  year = {2024},
  url = {https://yocharge.com/blog/electric-vehicle-incentives-by-country/},
  note = {Accessed October 21, 2025}
}

@article{bass1982note,
  title={A Note on Optimal Strategic Pricing of Technological Innovations},
  author={Bass, Frank M. and Bultez, Alain V.},
  journal={Marketing Science},
  volume={1},
  number={4},
  pages={371--378},
  year={1982},
  publisher={INFORMS},
  doi*={10.1287/mksc.1.4.371}
}

@article{kalish1983monopolist,
  title={Monopolist Pricing with Dynamic Demand and Production Cost},
  author={Kalish, Shlomo},
  journal={Marketing Science},
  volume={2},
  number={2},
  pages={135--159},
  year={1983},
  publisher={INFORMS},
  doi*={10.1287/mksc.2.2.135}
}

@article{kalish1983subsidy,
  title={Optimal Price Subsidy Policy for Accelerating the Diffusion of Innovation},
  author={Kalish, Shlomo and Lilien, Gary L.},
  journal={Marketing Science},
  volume={2},
  number={4},
  pages={407--420},
  year={1983},
  publisher={INFORMS},
  doi*={10.1287/mksc.2.4.407}
}

@article{thompson1984oligopoly,
  title={Optimal Pricing and Advertising Policies for New Product Oligopoly Models},
  author={Thompson, Gerald L. and Teng, Jinn-Tsair},
  journal={Marketing Science},
  volume={3},
  number={2},
  pages={148--168},
  year={1984},
  publisher={INFORMS},
  doi*={10.1287/mksc.3.2.148}
}

@article{mahajan1990new,
  title={New Product Diffusion Models in Marketing: A Review and Directions for Research},
  author={Mahajan, Vijay and Muller, Eitan and Bass, Frank M.},
  journal={Journal of Marketing},
  volume={54},
  number={1},
  pages={1--26},
  year={1990},
  publisher={American Marketing Association},
  doi*={10.2307/1252170}
}

@article{ryan1950acceptance,
  title={Acceptance and Diffusion of Hybrid Corn Seed in Two Iowa Communities},
  author={Ryan, Bryce and Gross, Neal C.},
  journal={Research Bulletin, Iowa Agriculture and Home Economics Experiment Station},
  volume={29},
  number={372},
  pages={663--708},
  year={1950},
  publisher={Iowa State College of Agriculture and Mechanic Arts},
  url* ={https://didawikinf.di.unipi.it/lib/exe/fetch.php/wma/agricultural_research_bulletin-v029-b372.pdf}
}

@article{krishnan1999optimal,
  title={Optimal Pricing Strategy for New Products},
  author={Krishnan, Trichy V. and Bass, Frank M. and Jain, Dipak C.},
  journal={Marketing Science},
  volume={18},
  number={3},
  pages={358--364},
  year={1999},
  publisher={INFORMS},
  doi*={10.1287/mksc.18.3.358}
}

@article{bayus1992dynamic,
  title={The Dynamic Pricing of Next Generation Consumer Durables},
  author={Bayus, Barry L.},
  journal={Marketing Science},
  volume={11},
  number={3},
  pages={251--265},
  year={1992},
  publisher={INFORMS},
  doi*={10.1287/mksc.11.3.251}
}

@article{dockner1988oligopoly,
  title={Optimal Pricing Strategies for New Products in Dynamic Oligopolies},
  author={Dockner, Engelbert and Jørgensen, Steffen},
  journal={Marketing Science},
  volume={7},
  number={4},
  pages={315--334},
  year={1988},
  publisher={INFORMS},
  doi*={10.1287/mksc.7.4.315}
}

@article{peres2010innovation,
  title={Innovation diffusion and new product growth models: A critical review and research directions},
  author={Peres, Renana and Muller, Eitan and Mahajan, Vijay},
  journal={International Journal of Research in Marketing},
  volume={27},
  number={2},
  pages={91--106},
  year={2010},
  publisher={Elsevier},
  doi*={10.1016/j.ijresmar.2009.12.012}
}

@article{sultan1990diffusion,
  title={A Meta-Analysis of Applications of Diffusion Models},
  author={Sultan, Fareena and Farley, John U. and Lehmann, Donald R.},
  journal={Journal of Marketing Research},
  volume={27},
  number={1},
  pages={70--77},
  year={1990},
  publisher={American Marketing Association},
  doi*={10.2307/3172552}
}

@misc{anylogic,
  author       = {AnyLogic},
  title        = {System Dynamics Tutorial: Bass Diffusion Model},
  year         = {2025},
  howpublished = {\url{https://anylogic.help/tutorials/system-dynamics/index.html}},
  note         = {Accessed: 2025-10-14}
}

@article{solar_rebate,
title = {The changing effectiveness of financial incentives: Theory and evidence from residential solar rebate programs in California},
journal = {Energy Policy},
volume = {162},
*pages = {112804},
year = {2022},
*ISSN* = {0301-4215},
*doi = {https://doi.org/10.1016/j.enpol.2022.112804},
*url = {https://www.sciencedirect.com/science/article/pii/S0301421522000295},
author = {Bixuan Sun and Ashwini Sankar},
keywords = {Solar PV, Financial incentives, Rebate programs, Program impact},
abstract = {Financial incentives, such as rebate programs, are widely used to promote the adoption of residential solar photovoltaic systems. This paper studies the changing effectiveness of rebate programs as the solar market evolves. We develop a theoretical model to first characterize individual treatment effects then aggregate them to regional effect, and hypothesize that the aggregate treatment effect of rebates first increases and then decreases with declining costs and increasing net metering rate. We empirically test our hypotheses by estimating the treatment effect of rebate rate in California from 2006 to 2017 and its interaction terms with installation cost and residential electricity rate. Results confirm our hypotheses and identify an installation cost of $8.84/W and a residential electricity rate of 22.36c/kWh as the turning points where rebate impacts start to decline. Our findings suggest that front-loaded rebate rate design can be an effective and efficient tool at promoting residential solar adoption.}
}

@article{peer_effects,
author = {Bollinger, Bryan and Gillingham, Kenneth},
title = {Peer Effects in the Diffusion of Solar Photovoltaic Panels},
journal = {Marketing Science},
volume = {31},
number = {6},
pages = {900-912},
year = {2012},
doi* = {10.1287/mksc.1120.0727},

URL* = { 
    
        https://doi.org/10.1287/mksc.1120.0727
    
    

},
eprint* = { 
    
        https://doi.org/10.1287/mksc.1120.0727
    
    

}
,
    abstract = { Social interaction (peer) effects are recognized as a potentially important factor in the diffusion of new products. In the case of environmentally friendly goods or technologies, both marketers and policy makers are interested in the presence of causal peer effects as social spillovers can be used to expedite adoption. We provide a methodology for the simple, straightforward identification of peer effects with sufficiently rich data, avoiding the biases that occur with traditional fixed effects estimation when using the past installed base of consumers in the reference group. We study the diffusion of solar photovoltaic panels in California and find that at the average number of owner-occupied homes in a zip code, an additional installation increases the probability of an adoption in the zip code by 0.78 percentage points. Our results provide valuable guidance to marketers designing strategies to increase referrals and reduce customer acquisition costs. They also provide insights into the diffusion process of environmentally friendly technologies. }
}

@article{bass,
 *ISSN* = {00251909, 15265501},
 *URL = {http://www.jstor.org/stable/2628128},
 abstract = {A growth model for the timing of initial purchase of new products is developed and tested empirically against data for eleven consumer durables. The basic assumption of the model is that the timing of a consumer's initial purchase is related to the number of previous buyers. A behavioral rationale for the model is offered in terms of innovative and imitative behavior. The model yields good predictions of the sales peak and the timing of the peak when applied to historical data. A long-range forecast is developed for the sales of color television sets.},
 author = {Frank M. Bass},
 journal = {Management Science},
 number = {5},
 pages = {215--227},
 publisher = {INFORMS},
 title = {A New Product Growth for Model Consumer Durables},
 urldate = {2023-10-23},
 volume = {15},
 year = {1969}
}

@article{gbm,
 ISSN* = {07322399, 1526548X},
 URL* = {http://www.jstor.org/stable/183674},
 abstract = {Over a large number of new products and technological innovations, the Bass diffusion model (Bass 1969) describes the empirical adoption curve quite well. In this study, we generalize the Bass model to include decision variables such as price and advertising. The generalized model reduces to the Bass model as a special case and explains why the Bass model works so well without including decision variables. We compare our generalized Bass model to other approaches from the literature for including decision variables into diffusion models, and our results provide both theoretical and empirical support for the generalized Bass model. We also show how our generalized Bass model can be used for product planning purposes.},
 author = {Frank M. Bass and Trichy V. Krishnan and Dipak C. Jain},
 journal = {Marketing Science},
 number = {3},
 pages = {203--223},
 publisher = {INFORMS},
 title = {Why the Bass Model Fits without Decision Variables},
 urldate = {2023-11-11},
 volume = {13},
 year = {1994}
}

@book{diffusion_innovations,
  title     = "Diffusion of Innovations",
  author    = "Rogers, E. M.",
  year      = 1962,
  publisher = "The Free Press",
  address   = "New York, NY"
}

@article{ddcm,
  title={Diffusion of Residential Solar Power Systems: A Dynamic Discrete Choice Approach},
  author={Souyris, Sebastian and Duan, Jason A and Balakrishnan, Anant and Rai, Varun},
  journal={Available at SSRN 4301666},
  year={2022}
}

@article{utility,
author = {Cosguner, Koray and Seetharaman, P. B. (Seethu)},
title = {Dynamic Pricing for New Products Using a Utility-Based Generalization of the Bass Diffusion Model},
journal = {Management Science},
volume = {68},
number = {3},
pages = {1904-1922},
year = {2022},
doi* = {10.1287/mnsc.2021.4257},
URL* = { 
https://doi.org/10.1287/mnsc.2021.4257
},
eprint* = { 
    
        https://doi.org/10.1287/mnsc.2021.4257
}
,
    abstract = { The Bass Model (BM) has an excellent track record in the realm of new product sales forecasting. However, its use for optimal dynamic pricing or advertising is relatively limited because the Generalized Bass Model (GBM), which extends the BM to handle marketing variables, uses only percentage changes in marketing variables, rather than their actual values. This restricts the GBM’s prescriptive use, for example, to derive the optimal price path for a new product, conditional on an assumed launch price, but not the launch price itself. In this paper, we employ a utility-based extension of the BM, which can yield normative prescriptions regarding both the introductory price and the price path after launch, for the new product. We offer two versions of this utility-based diffusion model, namely, the Bass-Gumbel Diffusion Model (BGDM) and the Bass-Logit Diffusion Model (BLDM), the latter of which has been previously used. We show that both the BGDM and BLDM handily outperform the GBM in forecasting new product sales using empirical data from four product categories. We discuss how to estimate the BGDM and BLDM in the absence of past sales data. We compare the optimal pricing policy of the BLDM with the GBM and derive optimal pricing policies that are implied by the BLDM under various ranges of model parameters. We illustrate a dynamic pricing approach that allows managers to derive optimal marketing policies in a computationally convenient manner and extend this approach to a competitive, multiproduct case.This paper was accepted by Gui Liberali, Management Science Special Section on Data-Driven Prescriptive Analytics. }
}

@misc{percentage_tax_credit,
  title         = "Making Our Homes More Efficient: Clean Energy Tax Credits for Consumers",
  author  = {{Office of Policy}},
  month         = jul,
  year          = "2024",
    url = "https://www.energy.gov/policy/articles/making-our-homes-more-efficient-clean-energy-tax-credits-consumers"
}

@misc{illinois_solar_for_all,
  title         = "Illinois Solar for All",
  author  = {{Illinois Power Agency}},
  month         = oct,
  year          = "2022",
    url = "https://ipa.illinois.gov/content/dam/soi/en/web/ipa/documents/10-17-22-ipa-il-solar-for-all-factsheet.pdf"
}

@misc{solar_data,
author = {{Lawrence Berkeley National Laboratory}},
  title         = "Tracking the Sun",
  organization  = "Lawrence Berkeley National Laboratory",
  month         = aug,
  year          = "2024",
  url = "https://emp.lbl.gov/tracking-the-sun"
}

@misc{census,
  title         = "U.S. Census Bureau QuickFacts",
  author  = {{United States Census Bureau}},
  year          = "2023",
  url = "https://www.census.gov/quickfacts/"
}

@misc{inflation_calculator,
  title         = "CPI Inflation Calculator",
  author  = {{Bureau of Labor Statistics}},
  year          = "2024",
  url = "https://data.bls.gov/cgi-bin/cpicalc.pl"
}

@article{dynamic_price_model,
 ISSN* = {00251909, 15265501},
 URL* = {http://www.jstor.org/stable/2629953},
 abstract = {The major points established in this paper are: classic marginal pricing is far from optimum for a rapidly evolving business; more appropriate dynamic models can be formulated if one has some feeling for the dominant evolutionary forces in the business environment; and, planning based on the dynamic models can lead to a significant improvement in the long run profit performance. Two developments in the management science literature, the experience curve phenomenon and market-penetration models, are used to illustrate the nature of the dynamic feedback between market and production activity which causes a new growth business to evolve. A specific illustrative example is offered which demonstrates the fact that dynamic price models can be used to test the long run consequences of specific pricing rules or to determine the optimum long run pricing scenario within the context of any constraints which a manager might wish to impose. As opposed to the conventional static theory which emphasizes the instantaneous profit flow, the dynamic models use an appropriately discounted accumulated profit as the major parameter for making value judgments. The specific example considered emphasizes the importance of these ideas for a growth market and suggests that dynamic models can lead to as much as an order of magnitude more profit in the long run than the conventional static theory. More modest, but significant, improvements in long run performance can be obtained in a moderate growth business.},
 author = {Bruce Robinson and Chet Lakhani},
 journal = {Management Science},
 number = {10},
 pages = {1113--1122},
 publisher = {INFORMS},
 title = {Dynamic Price Models for New-Product Planning},
 urldate = {2024-03-19},
 volume = {21},
 year = {1975}
}

@book{dp,
author = {Bertsekas, Dimitri P.},
title = {Dynamic Programming and Optimal Control, Vol. II},
year = {2007},
isbn* = {1886529302},
publisher = {Athena Scientific},
edition = {3rd},
abstract = {A major revision of the second volume of a textbook on the far-ranging algorithmic methododogy of Dynamic Programming, which can be used for optimal control, Markovian decision problems, planning and sequential decision making under uncertainty, and discrete/combinatorial optimization. The second volume is oriented towards mathematical analysis and computation, and treats infinite horizon problems extensively. New features of the 3rd edition are: 1) A major enlargement in size and scope: the length has increased by more than 50\%, and most of the old material has been restructured and/or revised. 2) Extensive coverage (more than 100 pages) of recent research on simulation-based approximate dynamic programming (neuro-dynamic programming), which allow the practical application of dynamic programming to large and complex problems. 3) An in-depth development of the average cost problem (more than 100 pages), including a full analysis of multichain problems, and an extensive analysis of infinite-spaces problems. 4) An introduction to infinite state space stochastic shortest path problems. 5) Expansion of the theory and use of contraction mappings in infinite state space problems and in neuro-dynamic programming. 6) A substantive appendix on the mathematical measure-theoretic issues that must be addressed for a rigorous theory of stochastic dynamic programming. Much supplementary material can be found in the book's web page: http://www.athenasc.com/dpbook.html}
}

@inbook{ift,
    author    = "B. M. Makarov and A. N. Podkorytov",
    title     = "Smooth Functions and Maps",
    publisher = "Springer",
    series   = "Moscow Lectures",
    chapter = "I",
    volume = "7",
    address  = "Moscow",
    year      = "2021"
}

@misc{ev_prices_incentives,
  title         = "Incentives and lower costs drive electric vehicle adoption in our Annual Energy Outlook",
  author  = {{U.S. Energy Information Administration}},
  year          = "2023",
  url = "https://www.eia.gov/todayinenergy/detail.php?id=56480"
}

@article{hbr_1976,
Abstract = {HBR first published this article in November 1950 as a practical guide to the problems involved in pricing new products. Particularly in the early stages of competition, it is necessary to estimate demand, anticipate the effect of various possible combinations of prices, and choose the most suitable promotion policy. Then as the product's market status matures, policy revisions become necessary, Joel Dean outlines the possible price strategies for each stage of a product's market evolution and the various grounds for making a choice. To update his original statement, Mr. Dean has written a retrospective comment, which appears at the end of this article. He amplifies his earlier article with insights from intervening years and in light of such developments as inflation. [ABSTRACT FROM AUTHOR]},
Author = {Dean, Joel},
ISSN* = {00178012},
Journal = {Harvard Business Review},
Keywords = {Pricing, New product development, Economic demand, Sales promotion, Product management, Product life cycle, Price inflation, Business planning, Business literature, Life cycle costing, Published reprints, Competition},
Number = {6},
Pages = {141 - 153},
Title = {Pricing policies for new products.},
Volume = {54},
URL* = {https://proxy2.library.illinois.edu/login?url=https://search.ebscohost.com/login.aspx?direct=true&db=bsu&AN=3867420&site=eds-live&scope=site},
Year = {1976},
}

@article{price_path,
author={Horsky,Dan},
year={1990},
month={Fall},
title={A DIFFUSION MODEL INCORPORATING PRODUCT BENEFITS, PRICE, INCOME AND INFORMATION},
journal={Marketing Science (1986-1998)},
volume={9},
number={4},
pages={342},
isbn={07322399},
language={en},
url*={https://www.proquest.com/scholarly-journals/diffusion-model-incorporating-product-benefits/docview/207344877/se-2},
}

@article{vandenbulte,
 ISSN* = {07322399, 1526548X},
 URL* = {http://www.jstor.org/stable/184230},
 author = {Van den Bulte, Christophe and Lilien, Gary L.},
 journal = {Marketing Science},
 number = {4},
 pages = {338--353},
 publisher = {INFORMS},
 title = {Bias and Systematic Change in the Parameter Estimates of Macro-Level Diffusion Models},
 urldate = {2025-10-20},
 volume = {16},
 year = {1997}
}

@article{krishnan2006,
author={Krishnan,Trichy V. and Jain,Dipak C.},
year={2006},
month={12},
title={Optimal Dynamic Advertising Policy for New Products},
journal={Management Science},
volume={52},
number={12},
pages={1957-1969},
isbn*={00251909},
language={English},
url*={https://www.proquest.com/scholarly-journals/optimal-dynamic-advertising-policy-new-products/docview/213259781/se-2},
}

\end{document}